# Influence of the Dufour effect on convection in binary gas mixtures

St. Hollinger and M. Lücke
*Institut für Theoretische Physik, Universität des Saarlandes*
*Postfach 151150, D 66041 Saarbrücken, Germany*

Linear and nonlinear properties of convection in binary fluid layers heated from below are investigated, in particular for gas parameters. A Galerkin approximation for realistic boundary conditions that describes stationary and oscillatory convection in the form of straight parallel rolls is used to determine the influence of the Dufour effect on the bifurcation behaviour of convective flow intensity, vertical heat current, and concentration mixing. The Dufour–induced changes in the bifurcation topology and the existence regimes of stationary and traveling wave convection are elucidated. To check the validity of the Galerkin results we compare with finite–difference numerical simulations of the full hydrodynamical field equations. Furthermore, we report on the scaling behaviour of linear properties of the stationary instability.

## I. INTRODUCTION

Convection in binary fluid mixtures heated from below [1,2] is described by balance equations for mass, momentum, heat, and concentration. The diffusive currents of heat and concentration that enter into the two latter balances are driven by generalized thermodynamic forces according to linear Onsager relations. They give rise to the Soret effect — temperature gradients change concentration — and to the Dufour efffect — concentration gradients change temperature. In binary *liquid* mixtures such as alcohol–water [3-10] or $^3$He–$^4$He [11-13] the Dufour effect is negligible. Most of the research activity in the field of convection in binary fluid mixtures has been focussed on these binary liquid mixtures.

However, in binary *gas* mixtures the Dufour effect is so large that it typically dominates the convective behaviour whenever the magnitude of the Soret coupling strength, i.e., of the separation ratio $\psi$ [2] is not negligible small. The importance of the Dufour effect in *gas* mixtures has two causes: (i) The Lewis number $L = D/\kappa$, i.e., the ratio of concentration diffusion constant $D$ and thermal diffusivity being of order 1 in *gas* mixtures is about 100 times larger than in *liquid* mixtures. (ii) The Dufour number $Q$ measuring the contribution to the generalized thermodynamic forces in the linear Onsager relations from gradients of the chemical potential that are caused by concentration gradients can be estimated [14] to be $Q \simeq 20$–$40$ in *gas* mixtures. Now, the Dufour effect changes the (dimensionless) equation of motion

$$(\partial_t + \mathbf{u} \cdot \boldsymbol{\nabla})T = (1 + QL\psi^2)\nabla^2 T - QL\psi \nabla^2 C \quad (1.1)$$

of the temperature field $T$ in two ways [14]. The "diagonal" term $QL\psi^2 \nabla^2 T$ reflects an enhancement of temperature diffusion of relative size $QL\psi^2$. The "offdiagonal" contribution $-QL\psi\nabla^2 C$ describes the direct effect of gradients in the concentration field $C$ on the temperature field. Both contributions to (1.1) are large when the size of the Soret coupling $\psi$ is not too small.

The influence of the Dufour effect on the onset behaviour of convection in binary mixtures was determined within a *linear* analysis [14] of the convective perturbations of the quiescent conductive state. Here we first extend the exact analytical linear results of Lee, Lucas, and Tyler [15] for the stationary instability. Then, we mainly investigate various *nonlinear* convective properties and how they are influenced by the Dufour effect — in particular for *gas* parameters. We mostly use an eight–mode Galerkin approximation to describe convection in the form of straight parallel rolls subject to realistic horizontal boundary conditions.

Similar models for binary *liquid* mixtures have to cope with two difficulties: Boundary layer phenomena caused by the smallness of the Lewis number, $L = \mathcal{O}(10^{-2})$, in *liquids* and the peculiar structure of the concentration field in traveling wave (TW) convection [16]. The restricted spatial resolution of a few–mode Galerkin truncation does not capture details of too fine a spatial field structure. Binary *gas* mixtures, on the other hand, are more favourable for such models: With $L = \mathcal{O}(1)$ concentration boundary layer problems are less severe and the existence range of TW solutions is significantly reduced in parameter space since not only $L$ but also the Prandtl number $\sigma$ is of order 1. In any case, we checked our analytical Galerkin results against finite–difference numerical test calculations of the full hydrodynamical field equations in order to assess the validity of the eight–mode Galerkin model. A positive feature of the latter is of course that it allows a convenient analysis of variations with the control parameters Rayleigh number and separation ratio $\psi$ and with the material parameters $L$, $\sigma$, $Q$.

The paper is organized as follows: In Sec.II the system, the equations, the boundary conditions, and the order parameters are described. Sec.III is preoccupied with analytical results of the stationary stability analysis of the quiescent heat conducting state. In Sec.IV we derive the Galerkin model and investigate the influence of the Dufour effect on linear properties of convective perturbations and on the nonlinear solutions of stationary and oscillatory convection. Sec.V contains comparisons of the model with linear and nonlinear results obtained from linear stability analyses and finite–difference numerical simulations of the full hydrodynamical field equations. Sec.VI summarizes our results. In an appendix we present the corrected version of the linear stability analysis [14] for more idealized free–slip, impermeable boundary conditions.

## II. SYSTEM

We consider a fluid layer of height $d$ between impervious, perfectly heat conducting horizontal plates which is exposed to a homogeneous vertical gravitational acceleration $g$ in $z$–direction. We impose a vertical temperature gradient so that the plates at $z = \mp \frac{d}{2}$ are kept at temperatures $T_0 \pm \frac{\Delta T}{2}$, where $T_0$ is the mean temperature of the fluid. The associated Rayleigh number

$$R = \frac{\alpha g d^3}{\nu \kappa} \Delta T \qquad (2.1)$$

is given by the thermal diffusivity $\kappa$, the kinematic viscosity $\nu$, and the thermal expansion coefficient

$$\alpha = -\frac{1}{\rho} \frac{\partial \rho(T, \hat{p}, C)}{\partial T} \quad . \qquad (2.2)$$

The solutal expansion coefficient is given by

$$\beta = -\frac{1}{\rho} \frac{\partial \rho(T, \hat{p}, C)}{\partial C} \quad , \qquad (2.3)$$

where $C$ denotes the concentration, $\hat{p}$ the pressure, and $\rho$ the fluid's density. Throughout most of this paper we use dimensionless units which scale lengths by $d$, times by $\frac{d^2}{\kappa}$, temperatures by $\frac{\nu \kappa}{\alpha g d^3}$ and concentrations by $\frac{\nu \kappa}{\beta g d^3}$.

### A. Equations

The hydrodynamic field equations governing the system's dynamics are well known [14]. In Oberbeck–Boussinesq approximation they are

$$\boldsymbol{\nabla} \cdot \hat{\mathbf{u}} = 0 \qquad (2.4a)$$

$$(\partial_t + \hat{\mathbf{u}} \cdot \boldsymbol{\nabla}) \hat{\mathbf{u}} = -\boldsymbol{\nabla} \left( \frac{\hat{p}}{\rho_0} + g z \right) + \nu \nabla^2 \hat{\mathbf{u}}$$
$$+ g \left[ \alpha (T - T_0) + \beta (C - C_0) \right] \mathbf{e}_z \qquad (2.4b)$$

$$(\partial_t + \hat{\mathbf{u}} \cdot \boldsymbol{\nabla}) C = D \nabla^2 C + D \frac{k_T}{T_0} \nabla^2 T \qquad (2.4c)$$

$$(\partial_t + \hat{\mathbf{u}} \cdot \boldsymbol{\nabla}) T = \kappa (1 + k_T^2 L a) \nabla^2 T$$
$$+ \kappa k_T L a T_0 \nabla^2 C \quad . \qquad (2.4d)$$

Here $C_0$ is the mean concentration of the mixture and $D$ is the concentration diffusion coefficient. The Lewis number $L = \frac{D}{\kappa}$ gives the ratio of time scales for concentration and heat diffusion. The Soret effect enters via the thermodiffusivity $k_T$ while

$$a = \frac{1}{c_p T_0} \frac{\partial \mu(T, \hat{p}, C)}{\partial C} \qquad (2.5)$$

quantifies the strength of the Dufour effect. In eq.(2.5) $c_p$ is the isobaric specific heat capacity and $\mu$ the chemical potential.

### B. Dimensionless deviations from the conductive state

The stationary solution of the Oberbeck–Boussinesq equations (OBE) describing the state of pure heat conduction without convection is

$$T_{cond} = T_0 - \frac{\Delta T}{d} z \qquad (2.6a)$$

$$C_{cond} = C_0 + \frac{k_T}{T_0} \frac{\Delta T}{d} z \qquad (2.6b)$$

$$\hat{p}_{cond} = \hat{p}(z=0) - \rho g z \left[ 1 + \left( \alpha - \beta \frac{k_T}{T_0} \right) \frac{\Delta T}{2d} z \right] \qquad (2.6c)$$

$$\hat{\mathbf{u}}_{cond} = 0 \quad . \qquad (2.6d)$$

Here $C_0$ is the mean concentration of the mixture. We pass over to reduced fields for the deviations from the conductive state

$$\theta = \frac{\alpha g d^3}{\nu \kappa} (T - T_{cond}) \qquad (2.7a)$$

$$c = \frac{\beta g d^3}{\nu \kappa} (C - C_{cond}) \qquad (2.7b)$$

$$p = \frac{d^2}{\rho_0 \kappa^2} (\hat{p} - \hat{p}_{cond}) \qquad (2.7c)$$

$$\mathbf{u} = \frac{d}{\kappa} \hat{\mathbf{u}} = (u, v, w) \qquad (2.7d)$$

that obey the equations

$$(\partial_t - \sigma \nabla^2) \nabla^2 w = \sigma (\partial_x^2 + \partial_y^2)(\theta + c) + NLT \qquad (2.8a)$$

$$(\partial_t + \mathbf{u} \cdot \boldsymbol{\nabla}) \theta = R w + (1 + L Q \psi^2) \nabla^2 \theta$$
$$- L Q \psi \nabla^2 c \qquad (2.8b)$$

$$(\partial_t + \mathbf{u} \cdot \boldsymbol{\nabla}) c = R \psi w + L \nabla^2 (c - \psi \theta) \qquad (2.8c)$$

$$\boldsymbol{\nabla} \cdot \mathbf{u} = 0 \qquad (2.8d)$$

$$NLT = \mathbf{e}_z \cdot [\boldsymbol{\nabla} \times \boldsymbol{\nabla} \times ((\mathbf{u} \cdot \boldsymbol{\nabla}) \mathbf{u})] \quad . \qquad (2.8e)$$

To derive (2.8a) we have applied twice the curl operator on eq.(2.4b). We have introduced the Prandtl number $\sigma = \frac{\nu}{\kappa}$. The separation ratio

$$\psi = -\frac{k_T}{T_0} \frac{\beta}{\alpha}$$

measures the Soret coupling. The Dufour number is

$$Q = \left(\frac{\alpha T_0}{\beta}\right)^2 a \quad .$$

The Dufour effect enters into (2.8b) diagonally via the term $LQ\psi^2\nabla^2\theta$ that reflects an enhancement of heat diffusion — $Q$ is positive — and offdiagonally via $-LQ\psi\nabla^2 c$ which represents concentration–induced changes in the temperature field. Thus for small Soret coupling $\psi$ we can expect only small Dufour effects on linear and nonlinear properties.

The parameters $L$, $\sigma$, $\psi$, and $Q$ depend on the mean temperature, concentration, and pressure of the fluid. Nevertheless, it is appropriate to characterize convective properties by $L$, $\sigma$, $\psi$, and $Q$ instead of by the three thermal equilibrium quantities. Note, furthermore, that $Q$ is known only poorly [17]. In order to select the range of parameters to be investigated here we used the following pieces of information: Hort et al. [14] have calculated Dufour numbers of order 10 using an ideal gas model, see also [17]. We limit ourselves mainly to the interval $(0, 20)$. Since in gases concentration, heat, and momentum diffuse on the same time scales we mostly investigate mixtures with Lewis and Prandtl number 1. The separation ratio is considered in the interval $(-1, 0.25)$, that can be expected to contain the experimentally accessible range.

### C. Boundary conditions

For a complete solution of the governing equations we need a set of boundary conditions for the three relevant fields $w$, $\theta$, $c$. Except for the Appendix we impose realistic no slip, impermeable (NSI) boundary conditions. Since the concentration flux at the no slip boundaries is purely diffusive,

$$\mathbf{J}_c = -L\boldsymbol{\nabla}(C - \psi T) \quad \text{for} \quad z = \pm\frac{1}{2} \quad , \tag{2.9}$$

we have to set

$$\partial_z(c - \psi\theta) = 0 \quad \text{for } z = \pm\frac{1}{2} \tag{2.10}$$

in order to avoid a vertical flux of solvent through the plates. In the conductive state the concentration flux vanishes identically. It is useful to introduce the field

$$\zeta(x, y, z; t) = c(x, y, z; t) - \psi\theta(x, y, z; t) \tag{2.11}$$

instead of $c(x, y, z; t)$ with the simpler boundary condition

$$\partial_z\zeta = 0 \quad \text{for } z = \pm\frac{1}{2} \quad . \tag{2.12}$$

The no slip boundary condition is described by

$$w = 0 = \partial_z w \quad \text{for } z = \pm\frac{1}{2} \quad . \tag{2.13}$$

Finally, since the temperature is fixed at the plates the deviation of the temperature from its conductive profile has to be zero for perfect conductors

$$\theta = 0 \quad \text{for } z = \pm\frac{1}{2} \quad . \tag{2.14}$$

More idealized, free slip, impermeable (FSI) boundary conditions are described in the Appendix.

Using the $\zeta$–field we get the following system of partial differential equations governing the convection in binary fluid mixtures

$$(\partial_t - \sigma\nabla^2)\nabla^2 w = \sigma(\partial_x^2 + \partial_y^2)\left[(1 + \psi)\theta + \zeta\right] + NLT \tag{2.15a}$$

$$(\partial_t + \mathbf{u}\cdot\boldsymbol{\nabla})\theta = Rw + \nabla^2\theta - LQ\psi\nabla^2\zeta \tag{2.15b}$$

$$(\partial_t + \mathbf{u}\cdot\boldsymbol{\nabla})\zeta = \mathcal{L}\nabla^2\zeta - \psi\nabla^2\theta \tag{2.15c}$$

$$\boldsymbol{\nabla}\cdot\mathbf{u} = 0 \tag{2.15d}$$

$$NLT = \mathbf{e}_z\cdot\left[\boldsymbol{\nabla}\times\boldsymbol{\nabla}\times((\mathbf{u}\cdot\boldsymbol{\nabla})\mathbf{u})\right] \tag{2.15e}$$

with the boundary conditions

$$\begin{aligned} w &= 0 = \partial_z w \\ \theta &= 0 = \partial_z\zeta \end{aligned} \quad \text{for } z\pm = \frac{1}{2} \quad . \tag{2.16}$$

In eq.(2.15c) we have introduced an effective Lewis number

$$\mathcal{L} = L(1 + Q\psi^2) \quad . \tag{2.17}$$

Therefore the Dufour effect is switched off by cancelling the term $LQ\psi\nabla^2\zeta$ *and* replacing $\mathcal{L}$ by $L$.

### D. Order parameters

To describe convection we shall use different order parameters. (i) The maximal vertical flow velocity $w_{max}$ directly measures the convective amplitude. (ii) The Nusselt number

$$N = 1 - \frac{1}{R}\partial_z <\theta>_{x,y}|_{z=\pm 1/2} \tag{2.18}$$

is the total vertical heat current through the layer reduced by the conductive part $R$. Here the brackets imply a lateral average. To avoid the problem of determining the bulk heat current in the presence of a Dufour effect we evaluate for convenience the vertical heat current through the fluid layer right at $z = \pm\frac{1}{2}$: Not only heat advection but also any Dufour–induced contribution to the heat transport from vertical concentration currents vanishes at the NSI–plates. The reduced vertical heat current carried by convection alone, $N - 1$, measures the squared convective field amplitudes. (iii) Since $w_{max}$ and $N - 1$ do not characterize the concentration field we use the "mixing parameter" [16]

$$M = \sqrt{\frac{<(C - C_0)^2>_{x,y,z}}{<(C_{cond} - C_0)^2>_{x,y,z}}} \quad . \tag{2.19}$$

$M$ is the variance of the concentration field reduced by its value in the conductive state and thus characterizes the magnitude of concentration variations around the mean $C_0$. In a perfectly mixed fluid, where all concentration deviations from $C_0$ vanish, $M$ would be zero while in the conductive state with the Soret–induced concentration gradient $M$ is defined to be 1. So $1 - M$ is an order parameter for convective states that is zero in the conductive state and approaches 1 for convection with perfect mixing. Finally, (iv) propagating convection rolls are characterized by the oscillation frequency of the traveling convection wave.

## III. SCALING BEHAVIOUR OF STATIONARY STABILITY PROPERTIES

As we will see later on the stationary stability threshold of the quiescent heat conducting state is the smallest one nearly all over the parameter space that is relevant for binary gas mixtures. Therefore, we compile, review, and extend in this section the exact analytical stationary stability analysis of Lee, Lucas, and Tyler [15] that is based on the method of Chandrasekhar [18] (see also the paper of Gutkowicz–Krusin, Collins, and Ross [19], where, however, the diagonal contribution of the Dufour effect to the heat balance was ignored). Our analysis reveals an interesting scaling behaviour of stationary stability properties that allows to scale away the Dufour effect. In addition to [15] we determine here the exact eigenfunctions. Furthermore, we present an analytical calculation of the zero wave number instability, an expansion of the critical Rayleigh number and the critical wave number around it, and further discussions, especially concerning the Dufour effect.

Using

$$\widetilde{R} = RS \quad ; \quad S = \frac{\mathcal{L}(1+\psi) + \psi}{L} \quad ; \quad \tau = \left(\widetilde{R}/k^4\right)^{1/3} \quad (3.1)$$

and

$$q_0 = ik\sqrt{\tau - 1} \quad (3.2a)$$

$$q_1 = \frac{k}{\sqrt{2}} \left[ \sqrt{\sqrt{1+\tau+\tau^2}+1+\frac{\tau}{2}} + i\sqrt{\sqrt{1+\tau+\tau^2}-1-\frac{\tau}{2}} \right] \quad (3.2b)$$

the stationary eigenfunctions on the marginal stability curve have the form

$$\begin{pmatrix} w \\ \theta \\ \zeta \end{pmatrix} = \begin{pmatrix} \hat{w} & \hat{w}_0 & \hat{w}_1 \\ \hat{\theta} & \hat{\theta}_0 & \hat{\theta}_1 \\ \hat{\zeta} & \hat{\zeta}_0 & \hat{\zeta}_1 \end{pmatrix} \begin{pmatrix} \cosh kz \\ \cos q_0 z \\ \cosh q_1 z \end{pmatrix} \cos kx + c.c. \quad . \quad (3.3)$$

Here the coordinate system is such that the horizontal wave number of the perturbation fields is $(k_x, k_y) = (k, 0)$. The linear equations of motion yield the relations

$$\hat{w} = 0 \quad ; \quad \hat{\zeta} = -(1+\psi)\hat{\theta} \quad (3.4a)$$

$$\begin{pmatrix} \hat{\theta}_0 \\ \hat{\zeta}_0 \end{pmatrix} = \frac{R}{Lk^2\tau} \begin{pmatrix} \mathcal{L} \\ \psi \end{pmatrix} \hat{w}_0 \quad (3.4b)$$

$$\begin{pmatrix} \hat{\theta}_1 \\ \hat{\zeta}_1 \end{pmatrix} = -\frac{R}{Lk^2\tau} \cdot \frac{1-i\sqrt{3}}{2} \begin{pmatrix} \mathcal{L} \\ \psi \end{pmatrix} \hat{w}_1 \quad (3.4c)$$

between amplitudes so that four unknowns $\hat{\theta}$, $\hat{w}_0$, $\hat{w}_1$, and $\hat{w}_1^*$ ($\hat{\theta}$ and $\hat{w}_0$ are real) and four boundary conditions remain. A solvability criterium yields the relation

$$k\tanh\frac{k}{2}\left\{Im\left[\left(\sqrt{3}+i\right)q_1\tanh\frac{q_1}{2}\right] + q_0\tan\frac{q_0}{2}\right\} =$$
$$p\left\{q_0\tan\frac{q_0}{2} Im\left[\left(\sqrt{3}-i\right)q_1\tanh\frac{q_1}{2}\right] - |q_1\tanh\frac{q_1}{2}|^2\right\} \quad (3.5)$$

between $\tau$ and $k$. It determines the marginal stability curve $\widetilde{R}_{stab}(k;p)$ depending on the parameter $p$ given below. Then one can determine also the marginal amplitudes $\hat{w}_0$, $\hat{w}_1$, and $\hat{w}_1^*$ — with the normalization chosen such that $2(1+p)\hat{\zeta} = -\tau^2 k^2 p$, they depend only on $k$ and $p$. The nonlinear combination

$$p = \frac{\psi}{\mathcal{L}(1+\psi)} = \frac{\psi}{L(1+Q\psi^2)(1+\psi)} \quad (3.6)$$

of parameters $L$, $Q$, and $\psi$ that was called $H$ by Lee, Lucas, and Tyler [15] is a scaling variable. Note that $p$ vanishes in the pure–fluid–limit, $\psi = 0$, of vanishing Soret effect.

The marginal curve

$$\widetilde{R}_{stab}(k;p) = \tau^3(k;p)k^4 \quad (3.7)$$

obtained from solving (3.5) for $\tau$ depends only via $p$ on the parameters $L$, $Q$, and $\psi$. Therefore, the critical wave number $k_c = k_c(p)$ that solves $\partial \widetilde{R}_{stab}(k;p)/\partial k = 0$ is only a function of $p$. On the other hand, the critical Rayleigh number

$$R_c = \frac{1}{S}\widetilde{R}_{stab}(k_c(p);p) = \frac{1}{S}\widetilde{R}_c(p) \quad (3.8)$$

is a function of $p$ and of

$$S = \frac{\mathcal{L}(1+\psi) + \psi}{L} = \frac{\psi}{L}\frac{1+p}{p} \quad . \quad (3.9)$$

The scaling relation (3.8) with $\mathcal{L} = L(1+Q\psi^2)$

$$R_c(L, Q, \psi) = \frac{L}{\psi + \mathcal{L}(1+\psi)}\widetilde{R}_c\left(\frac{\psi}{\mathcal{L}(1+\psi)}\right) \quad (3.10)$$

implies a significant simplification for practical calculations of critical stationary properties: Only two functions, $k_c(p)$ and $\widetilde{R}_c(p)$, have to be determined as functions of $p$ to get $k_c$ and $R_c$ for all $L$, $Q$, and $\psi$ combinations. In Table I, we list, for several $p$, the scaled critical

Rayleigh numbers, $\widetilde{R}_c(p)$, and the critical wave numbers $k_c(p)$. Using $\tau_c = (\widetilde{R}_c/k_c^4)^{1/3}$ (3.1), one then can determine the critical vertical exponents $q_0$ and $q_1$ (3.2). The vertical profiles of the critical eigenfunctions (3.3) can be obtained from $\hat{w}_0$ and $\hat{w}_1$ in Table I. The amplitude $\hat{\zeta}$ appearing in (3.3, 3.4) is fixed by $2(1+p)\hat{\zeta} = -\tau_c^2 k_c^2 p$.

For $p = 0$, i.e., in the pure–fluid–limit $\psi = 0$ the scaling factor $S = 1$ and the scaling function has the critical value, $R_c^0$, of the one-component fluid:

$$\widetilde{R}_c(p{=}0) = R_c^0 = 1707.762 \quad \text{and} \quad k_c(p{=}0) = k_c^0 = 3.11633 \quad . \tag{3.11}$$

The marginal stability curve shows the scaling behaviour

$$R_{stab}(k; L, Q, \psi) = \frac{1}{S}\widetilde{R}_{stab}(k; p) \tag{3.12}$$

where $\widetilde{R}_{stab}$ (3.7) is defined by the solution of (3.5). Thus, the Dufour effect can be scaled away in stationary stability properties. Higher stability thresholds for odd eigenfunctions can be obtained by replacing cos by sin and cosh by sinh in the terms of (3.3) containing the vertical spatial dependence. This results in an equation of the same form as (3.5), however, with tanh replaced by coth and tan by $-$cot. The case of heating from above with $\tau < 1$ and imaginary $q_0$ can also be treated. Note that $q_0 \tan(q_0/2)$ as well as $q_0 \cot(q_0/2)$ are real for imaginary $q_0$.

In Fig. 1 we show (a) the critical reduced wave number $k_c(p)/k_c^0$ and (b) the reduced scaling function $\widetilde{R}_c(p)/R_c^0$, both versus p. For $p \to -1$, $k_c$ goes to about 7.48 and the scaling factor $S$ in (3.8) goes to zero, which is the reason for the divergence of $R_c$. At $p_0 = 131/34 \approx 3.85$ the critical wave number vanishes. We have evaluated $p_0$ and the stability behaviour in the neighbourhood of $p_0$ by a Taylor expansion of equation (3.5) up to order $k^{20/3}$ using the ansatz $\tau = k^{-4/3}(a + bk^2 + ck^4)$ that follows from expanding the scaled marginal Rayleigh number $\widetilde{R}_{stab}$ as

$$\widetilde{R}_{stab}(k) = \tau^3 k^4 = (a + bk^2 + ck^4 + \mathcal{O}(k^6))^3$$
$$= a^3 + 3a^2 bk^2 + 3a(b^2 + ac)k^4 + \mathcal{O}(k^6) \quad . \tag{3.13}$$

The value $p_0$ for which the minimum of $\widetilde{R}_{stab}$ is located at $k_c = 0$ is determined by the requirement $3a^2 b = 0$. In that case $\widetilde{R}_{stab}(k)$ increases proportional to $k^4$. Performing the expansion we get

$$R_{stab}\frac{\psi}{L} = \frac{p}{1+p}\widetilde{R}_{stab}$$
$$= 720 + \frac{2040}{77p}\left(p - \frac{131}{34}\right)k^2$$
$$+ \frac{1700115}{1310309 p^2}\left(p^2 + \frac{2033552}{340023}p + \frac{3779327}{340023}\right)k^4$$
$$+ \mathcal{O}(k^6) \quad . \tag{3.14}$$

For values of $p$ slightly below $p_0 = 131/34 = 3.85294...$, we can calculate

$$k_c^2 \approx \frac{3471468 \, p \left(\frac{131}{34} - p\right)}{340023 \, p^2 + 2033552 \, p + 3779327} \quad . \tag{3.15}$$

In Fig. 1 we show this $k_c$ (3.15) by a dashed line in comparison to the exact result. Gutkowicz-Krusin et al. [19] and Knobloch and Moore [20] have given the expansion of $R_{stab}\psi/L$ up to order $k^0$ thus obtaining the above stated value of 720. In (3.14) we present in addition the quadratic and quartic order and the $p$-dependence of $k_c$ whose determination uses the quartic order term in (3.14).

In the literature dealing with binary *liquid* mixtures there are several expressions for the separation ratio $\psi_0$ for which $k_c = 0$. For zero Dufour effect, our calculation leads to

$$\psi_0(Q = 0) = \frac{L}{f - L} \tag{3.16}$$

with $f = 1/p_0$. Linz et al. [21] and Lhost et al. [22] have obtained from Galerkin approximations using free-slip, impermeable (FSI) and NSI boundary conditions values of $f \approx 1.62$ and $f \approx 0.37$, respectively. Knobloch and Moore [20] have extracted $f \approx 0.26$ out of their numerical stability analysis which has to be compared with our exact result of $f = 34/131 = 0.25954....$

Now we discuss the influence of the Dufour effect on the critical wave number's $\psi$-dependence. Hort et al. [14] found, within their FSI Galerkin approximation, that with increasing $Q$ the curve $k_c(\psi)$ formed a saddle at $\psi = -1/3$ when $Q$ reached the value 27 for $L = 1$. This behaviour holds also for the exact result with NSI boundary conditions. To show this, consider $dk_c/d\psi = dp/d\psi \cdot dk_c/dp$. Since $k_c(p)$ (Fig. 1) is monotonous the extrema of $k_c(\psi)$ are given by the zeros of $dp/d\psi$, i.e., the roots of the third-order polynomial

$$\psi^3 + \frac{1}{2}\psi^2 - \frac{1}{2Q} = 0 \quad . \tag{3.17}$$

One root is always greater than zero and not of interest. Two additional real roots first occur at $Q = 27$, $\psi = -1/3$ causing a saddle in $k_c(\psi)$ there. The appearence of this saddle is independent of $L$ and $\sigma$ and holds for the exact NSI stability analysis as well as for the model [14].

We finally should like to mention that an application of the method presented here to the oscillatory stability analysis requires more numerical effort [20,23] because the vertical wave numbers $q$ are not simple third-order roots but solutions of a fourth-order polynomial depending also on frequency $\omega$. Therefore, exact analytical results for the oscillatory threshold, the critical wave number, and the Hopf frequency do not seem to be feasible.

## IV. GALERKIN APPROXIMATION

In this section we present our Galerkin model for realistic boundary conditions. Starting from this model we carry out a linear stability analysis of the conductive state before calculating nonlinear states, stationary as well as oscillatory ones.

### A. NSI mode truncation and model

To describe convection in the form of straight rolls we truncate the spatial mode expansion appropriate to NSI conditions by

$$w(x, z; t) = \left[ w_{11}(t)\, e^{-ikx} + c.c. \right] C_1(z) \quad (4.1a)$$

$$\theta(x, z; t) = \left[ \theta_{11}(t)\, e^{-ikx} + c.c. \right] \sqrt{2} \cos \pi z$$
$$+ \theta_{02}(t) \sqrt{2} \sin 2\pi z \quad (4.1b)$$

$$\zeta(x, z; t) = \left[ \zeta_{10}(t)\, e^{-ikx} + c.c. \right] + \zeta_{01}(t) \sqrt{2} \cos \pi z \quad (4.1c)$$

where $C_1(z) = \frac{\cosh \lambda_1 z}{\cosh \lambda_1 / 2} - \frac{\cos \lambda_1 z}{\cos \lambda_1 / 2}$ denotes the first even Chandrasekhar function with $\lambda_1 = 4.73004$. Since the modes $\theta_{01}$ and $\zeta_{02}$ are linearly damped within an enlarged model's frame, as we have calculated before restricting the mode truncation to the above form, we do not display them explicitly. Furthermore, these two modes would violate a mirror-glide symmetry of the $\theta$- and $\zeta$-field that was found [24] for stationary and traveling roll patterns. The no-slip boundary condition is guaranteed by using the Chandrasekhar function. Impermeability of the plates is ensured by confining the $\zeta$ ansatz to modes with vanishing $z$-derivatives at $z = \pm \frac{1}{2}$.

Projecting the OBE (2.15) onto the eight modes contained in our truncation (4.1), we obtain the following generalized Lorenz model

$$\tau \dot{\mathbf{X}} = -\widetilde{\sigma} \hat{f} \mathbf{X} + \widetilde{\sigma} \hat{g} \left[(1 + \psi) \mathbf{Y} + \alpha_1 \mathbf{U}\right] \quad (4.2a)$$

$$\tau \dot{\mathbf{Y}} = -\hat{q}^2 \mathbf{Y} + (r - Z) \mathbf{X} + \beta_1 \widetilde{\psi} \hat{k}^2 \mathbf{U} \quad (4.2b)$$

$$\tau \dot{\mathbf{U}} = -\beta_1 \mathcal{L} \hat{k}^2 \mathbf{U} + \hat{q}^2 \psi \mathbf{Y} + V \mathbf{X} \quad (4.2c)$$

$$\tau \dot{Z} = -b \left[ Z - \mathbf{X} \cdot \mathbf{Y} + \frac{\eta_1}{4 \mu_1} \widetilde{\psi} V \right] \quad (4.2d)$$

$$\tau \dot{V} = -\frac{b}{4} \left[\mu_1 \mathbf{X} \cdot \mathbf{U} + \eta_1 \psi Z + \mathcal{L} V \right] \quad (4.2e)$$

with

$$\widetilde{\psi} = \frac{8}{\pi^2} L Q \psi \quad , \quad \mathcal{L} = L(1 + Q \psi^2) \quad . \quad (4.3)$$

We used for the critical modes the following vector notation

$$\begin{aligned}
\mathbf{X} &= (X_1, X_2) = \tfrac{2\sqrt{2} a_1}{q_c^0}\, (Re\ w_{11}, Im\ w_{11}) \\
\mathbf{Y} &= (Y_1, Y_2) = 2 a_3 \tfrac{q_c^0}{R_c^0}\, (Re\ \theta_{11}, Im\ \theta_{11}) \quad (4.4) \\
\mathbf{U} &= (U_1, U_2) = a_3 \tfrac{\pi q_c^0}{\sqrt{2} R_c^0}\, (Re\ \zeta_{10}, Im\ \zeta_{10}) \quad .
\end{aligned}$$

These modes drive nonlinear ones via the eqs (4.2):

$$Z = 2\sqrt{2} a_3 \frac{\pi}{R_c^0} \theta_{02} \quad \text{and} \quad V = -\frac{\pi^2}{2\sqrt{2} R_c^0} \zeta_{01} \quad . \quad (4.5)$$

The constants are

$$\begin{aligned}
a_1 &= 2\pi \lambda_1^2 \left( \tfrac{1}{\lambda_1^4 - \pi^4} - \tfrac{3}{\lambda_1^4 - 81 \pi^4} \right) &= 0.4058 \\
a_2 &= \tfrac{4 \pi \lambda_1^2}{\lambda_1^4 - \pi^4} &= 0.6974 \\
a_3 &= \tfrac{a_1}{a_2} &= 0.5818 \\
a_4 &= 2\lambda_1 \tanh \tfrac{\lambda_1}{2} - \lambda_1^2 \tanh^2 \tfrac{\lambda_1}{2} &= -12.3026 \quad (4.6) \\
\alpha_1 &= \tfrac{8}{a_2 \pi \lambda_1} \tanh \tfrac{\lambda_1}{2} &= 0.7585 \\
\beta_1 &= \tfrac{k_c^{0 2}}{k_c^{0 2} + \pi^2} &= 0.4930 \\
\eta_1 &= \tfrac{4}{3 a_3} &= 2.2916 \\
\mu_1 &= \tfrac{1}{2 a_3^2} &= 1.4770
\end{aligned}$$

The quantities $k_c^0 = 3.098$, $q_c^{0 2} = k_c^{0 2} + \pi^2$, $R_c^0 = 1728.38$, $\tau = \frac{1}{q_c^{0 2}} = 0.05138$ and $b = \frac{4\pi^2}{q_c^{0 2}} = 2.0282$ are critical properties of the model for $\psi = 0$ [25]. $\widetilde{\sigma} = 1.943 \sigma$ denotes a rescaled Prandtl number caused by the no-slip mode truncation [25]. In addition we use the reduced Rayleigh number $r$ and the wave numbers $\hat{k}$ and $\hat{q}$ defined by

$$r = \frac{R}{R_c^0} \quad , \quad \hat{k} = \frac{k}{k_c^0} \quad , \quad \hat{q}^2 = \frac{q^2}{q_c^{0 2}} = \frac{k^2 + \pi^2}{k_c^{0 2} + \pi^2} \quad (4.7a)$$

as well as the quantities

$$\hat{f} = \frac{\lambda_1^4 + k^4 - 2 k^2 a_4}{k^2 - a_4} \frac{k_c^{0 2} - a_4}{\lambda_1^4 + k_c^{0 4} - 2 k_c^{0 2} a_4}$$

$$\text{and} \quad \hat{g} = \frac{k^2}{k^2 - a_4} \frac{k_c^{0 2} - a_4}{k_c^{0 2}} \quad (4.7b)$$

so that $\hat{k} = \hat{q} = \hat{f} = \hat{g} = 1$ for $k = k_c^0$. This notation is also used by Lhost et al. [22] who have examined the onset of convection in binary mixtures neglecting the Dufour effect and by Niederländer et al. [25] who have derived the analagous NSI model for one component fluids. Note, however, that in [25] the modes are reduced by $k$-dependent quantities — see, e.g., eqs. (2.8-2.10) in Ref. [25]. On the other hand, here we reduce the modes by critical properties of the $\psi = 0$ reference system. Thus, here all $k$-dependence of the system is displayed *explicitly* in the model equations (4.2).

The Dufour effect influences the mode balances in two distinct ways: It causes a driving of the temperature modes $\mathbf{Y}$ and $Z$ by the $\zeta$-field modes $\mathbf{U}$ and $V$, respectively. The associated coupling strength $\widetilde{\psi} = \frac{8}{\pi^2} L Q \psi$ vanishes when $Q = 0$. Furthermore, in the $\zeta$-field equations for $\mathbf{U}$ and $V$, the Lewis number $L$ is replaced by an effective one, $\mathcal{L} = L(1 + Q \psi^2)$.

To facilitate the quantitative comparison with experiments or numerical solutions of the full field equations we evaluate among others the order parameters defined in Sec. II D. Thus the maximal vertical flow velocity is given within the model by

$$w_{max} = \frac{q_c^0}{\sqrt{2}a_1} C_1(0) \, | \, \mathbf{X} \, | = 12.20 \, | \, \mathbf{X} \, | \quad (4.8)$$

in terms of the amplitude of mode $\mathbf{X}$. The reduced vertical convective heat current, evaluated at the plates, is

$$N - 1 = \frac{2\pi\sqrt{2}}{R}\theta_{02} = \frac{Z}{a_3 r} \quad . \quad (4.9)$$

Thus the Nusselt number is related to the mode $Z$. As an aside we mention here a deficiency of the no–slip Galerkin approximation that was discussed in more detail in [25]: Since the velocity field is expanded in Chandrasekhar functions, i.e., a nontrigonometric basis the stationary vertical heat current is not $z$–independent.

Into the mixing parameter $M$ (2.19) the temperature modes as well as the $\zeta$–field modes enter

$$M^2 = 1 + \frac{24}{r^2\psi^2}\frac{2}{a_3^2\pi^2 q_c^{0\,2}}\left[\mathbf{U}^2 + \frac{\pi^2}{8}\psi^2 \mathbf{Y}^2 + 2\psi\mathbf{Y}\cdot\mathbf{U}\right]$$
$$+ \frac{24}{r^2\psi^2}\frac{4}{\pi^2}\left[V^2 + \frac{\pi^2\psi^2}{64 a_3^2}Z^2 - \frac{2\psi}{3 a_3}ZV\right]$$
$$+ \frac{6}{r\psi\pi^4}\left\{32 V - \frac{\pi^2}{a_3}\psi Z\right\} \quad . \quad (4.10)$$

### B. Linear stability analysis

We start the discussion of our model with the investigation of the linear stability of the conductive fixed point where all mode amplitudes vanish. The nonlinear modes $Z$ and $V$ do not couple linearly into the equations for $\mathbf{X}$, $\mathbf{Y}$, $\mathbf{U}$ and are damped away. Therefore we have to seek the stability thresholds of the matrix system:

$$\tau \partial_t \begin{pmatrix} \mathbf{X} \\ \mathbf{Y} \\ \mathbf{U} \end{pmatrix} = \begin{pmatrix} -\tilde{\sigma}\hat{f} & \tilde{\sigma}\hat{g}(1+\psi) & \tilde{\sigma}\hat{g}\alpha_1 \\ r & -\hat{q}^2 & \beta_1\widetilde{\psi}\hat{k}^2 \\ 0 & \hat{q}^2\psi & -\beta_1\mathcal{L}\hat{k}^2 \end{pmatrix} \begin{pmatrix} \mathbf{X} \\ \mathbf{Y} \\ \mathbf{U} \end{pmatrix} \quad .$$
(4.11)

We calculate the stationary stability curve $r_{stat}(\hat{k})$ to be

$$r_{stat}(\hat{k}) \;=\; \frac{\hat{f}\hat{q}^2}{\hat{g}} \frac{1 - \dfrac{8Q\psi^2}{\pi^2(1+Q\psi^2)}}{1 + \psi\left(1 + \dfrac{\alpha_1}{\widehat{\mathcal{L}}}\right)} \quad . \quad (4.12)$$

The critical stationary wave number follows from

$$0 = \hat{k}_{stat}^{c\,6} + f_4\,\hat{k}_{stat}^{c\,4} + f_2\,\hat{k}_{stat}^{c\,2} + f_0 \quad (4.13)$$

with the coefficients

$$f_4 = \frac{\gamma_1 - \gamma_3}{2} + \frac{3\psi}{2\mathcal{L}}\frac{\gamma_1\alpha_1}{1+\psi\left(1+\dfrac{\alpha_1}{\mathcal{L}}\right)} \quad (4.14a)$$

$$f_2 = \frac{\psi}{\mathcal{L}}\frac{\gamma_1(\gamma_1-\gamma_3)\alpha_1}{1+\psi\left(1+\dfrac{\alpha_1}{\mathcal{L}}\right)} \quad (4.14b)$$

$$f_0 = \frac{\gamma_1\gamma_2}{2} + \frac{\psi}{2\mathcal{L}}\frac{\gamma_1\alpha_1(\gamma_2-\gamma_3\gamma_1)}{1+\psi\left(1+\dfrac{\alpha_1}{\mathcal{L}}\right)} \quad (4.14c)$$

and the numbers

$$\gamma_1 = \frac{\pi^2}{k_c^{0\,2}} = 1.029 \;;\; \gamma_2 = \frac{\lambda_1^4}{k_c^{0\,4}} = 5.437 \;;\; \gamma_3 = 2\frac{a_4}{k_c^{0\,2}} = -2.564 \;.$$
(4.14d)

The critical stationary wave number as a root of (4.13) is a function of the scaling parameter $p = \frac{\psi}{\mathcal{L}(1+\psi)}$ only — different $\psi$, $L$, $Q$ combinations for which $p$ is the same yield the same $k_{stat}^c(p)$. Such a scaling behaviour was found analytically in the exact stationary stability analysis of the full field equations by [15] and in Sec.III. However, the marginal curve $r_{stat}(\hat{k})$ (4.12) of the model does not show the full scaling behaviour seen in Sec.III.

The oscillatory stability curve is given by

$$r_{osc}(\hat{k}) \;=\; \frac{\hat{f}\hat{q}^2}{\hat{g}} \frac{\left[1+\widehat{\mathcal{L}}\right]\left[\left(1+\widehat{\widetilde{\sigma}}\right)\left(1+\dfrac{\widehat{\mathcal{L}}}{\widehat{\widetilde{\sigma}}}\right) - \dfrac{\psi\widehat{\widetilde{\psi}}}{\widehat{\widetilde{\sigma}}}\right]}{(1+\psi)(1+\widehat{\widetilde{\sigma}}) - \alpha_1\psi}$$
(4.15)

and the Hopf frequency at $r_{osc}(\hat{k})$

$$\omega_H^2 \frac{\tau^2}{\hat{q}^4} = -\widehat{\mathcal{L}}^2 - \alpha_1\psi\,\frac{(1+\widehat{\mathcal{L}})(\widehat{\widetilde{\sigma}}+\widehat{\mathcal{L}})}{(1+\psi)(1+\widehat{\widetilde{\sigma}}) - \alpha_1\psi}$$
$$+ \psi\widehat{\widetilde{\psi}}\,\frac{(1+\psi)(\widehat{\mathcal{L}}-\widehat{\widetilde{\sigma}})+\alpha_1\psi}{(1+\psi)(1+\widehat{\widetilde{\sigma}})-\alpha_1\psi} \quad . \quad (4.16)$$

Here we have introduced wave number dependent fluid parameters

$$\widehat{\widetilde{\sigma}} = \frac{\hat{f}}{\hat{q}^2}\widetilde{\sigma} \;,\; \widehat{\mathcal{L}} = \beta_1\frac{\hat{k}^2}{\hat{q}^2}\mathcal{L} \;,\; \text{and} \;\; \widehat{\widetilde{\psi}} = \beta_1\frac{\hat{k}^2}{\hat{q}^2}\widetilde{\psi} \;. \quad (4.17)$$

Since the square of the critical oscillatory wave number is given by a root of a polynomial of at least degree ten we evaluated $\hat{k}_{osc}^c$ numerically by minimizing $r_{osc}(\hat{k})$.

Fig. 2 shows the main results of our stability analysis: The Dufour effect destabilizes (stabilizes) the conductive state against the growth of stationary (oscillatory) convection and it shifts the critical curves $r_{stat}^c(\psi)$ and $r_{osc}^c(\psi)$ towards smaller $\psi$. The critical stationary wave number $k_{stat}^c(\psi)$ forms a saddle for $Q = 27$ and $\psi = -\frac{1}{3}$. The wave number $k_{osc}^c(\psi)$ of critical oscillatory patterns decreases and the difference, $k_{stat}^c - k_{osc}^c$, increases with increasing Dufour effect. The Hopf frequency decreases with growing $Q$. An important fact is that the $\psi$-range of oscillatory instability sharply shrinks with increasing Dufour number. It remains an experimental challenge to prepare mixtures that have the right $Q$-$\psi$ parameter combinations to see this behaviour. All the above described properties are the same as those obtained from the full field equations with a numerically performed shooting analysis (compare, e.g., Fig. 7 of Ref. [14] with our Fig. 2). We refer to Ref. [14] for a more detailed discussion of linear properties which are not the main topic of this paper.

Finally, we mention that the analytic oscillatory stability analysis presented in [14] for *idealized* FSI boundary conditions contains a mistake. In the Appendix we present the correct formulae and a figure showing these results. We find that our corrected FSI results are closer to the exact NSI curves than the FSI results of Ref. [14].

### C. Nonlinear convective states

Here we elucidate the influence of the Dufour effect on porperties such as strength of convection, bifurcation behaviour, heat flux, and concentration mixing in nonlinear states of stationary and oscillatory convection.

#### 1. Stationary convection

The stationary solutions of our model representing steady overturning convection (SOC) in the form of straight rolls are given by

$$\mathbf{X}^2 = -\frac{\alpha}{2} \pm \sqrt{\left(\frac{\alpha}{2}\right)^2 - \beta} \quad (4.18\text{a})$$

$$\mathbf{Y} = -F \left[\frac{\beta_1}{\mu_1}\mathcal{L}\hat{k}^2 \left(\mathcal{L} - \frac{\eta_1^2}{4\mu_1}\psi\widetilde{\psi}\right) + \mathbf{X}^2\right] \mathbf{X} \quad (4.18\text{b})$$

$$Z = -F \left[\frac{\beta_1}{\mu_1}\mathcal{L}^2\hat{k}^2 + \frac{\eta_1}{4\mu_1}\hat{q}^2\psi\widetilde{\psi} + \mathbf{X}^2\right] \mathbf{X}^2 \quad (4.18\text{c})$$

$$\mathbf{U} = \frac{\eta_1}{\mu_1}F\psi \left[-\frac{1}{\eta_1}\mathcal{L}\hat{q}^2 + \frac{\eta_1}{4\mu_1}\hat{q}^2\psi\widetilde{\psi} + \mathbf{X}^2\right]\mathbf{X} \quad (4.18\text{d})$$

$$V = F\psi \left[\frac{\beta_1\eta_1}{\mu_1}\mathcal{L}\hat{k}^2 + \hat{q}^2\right]\mathbf{X}^2 \quad (4.18\text{e})$$

where we have introduced the following abbreviations:

$$F^{-1} = \frac{\hat{g}}{\hat{f}}\left(\frac{\eta_1^2}{4\mu_1^2}\psi\widetilde{\psi} - \frac{\mathcal{L}}{\mu_1}\right)\left(\beta_1\hat{k}^2\mathcal{L}(1+\psi) + \alpha_1\hat{q}^2\psi\right)$$

$$-\frac{\hat{g}}{\hat{f}}\mathbf{X}^2\left[1 + \psi\left(1 - \frac{\alpha_1\eta_1}{\mu_1}\right)\right] \quad (4.18\text{f})$$

$$\alpha = \hat{q}^2 + \frac{\beta_1}{\mu_1}\mathcal{L}^2\hat{k}^2 + \frac{\eta_1\beta_1}{\mu_1}\psi\widetilde{\psi}(\hat{k}^2 + \frac{\hat{q}^2}{4\beta_1})$$

$$-r\frac{\hat{g}}{\hat{f}}\left[1 + \psi\left(1 - \frac{\alpha_1\eta_1}{\mu_1}\right)\right] \quad (4.18\text{g})$$

$$\beta = \frac{\beta_1}{\mu_1}\mathcal{L}^2\hat{k}^2\hat{q}^2 \left(1 - \frac{\eta_1^2}{4\mu_1}\frac{\psi\widetilde{\psi}}{\mathcal{L}}\right)\left(1 - \frac{\psi\widetilde{\psi}}{\mathcal{L}}\right)\left(1 - \frac{r}{r_{stat}}\right). \quad (4.18\text{h})$$

The formulae (4.18) for the SOC solution are structurally similar to those of the analogous models [26,21] without Dufour effect and derived for idealized FSP [26] or FSI [21] boundary conditions. Also here, like in [26,21], the SOC solution does not depend on the Prandtl number.

A positive $\mathbf{X}^2$ bifurcates according to (4.18a) out of the conductive solution, $\mathbf{X} = 0$, at the stationary threshold $r_{stat}$ where $\beta$ (4.18h) goes to zero. The influence of the Dufour effect on the convective intensity can be seen in Fig. 3. There we display $\mathbf{X}^2$, the square of the vertical velocity mode, versus $r$ for $\psi = -0.25$, $L = 1$, and $k = k_c^0$ for several values of the Dufour number $Q$. For a fixed Rayleigh number $r \gtrsim 2$ the strength of convection is reduced with increasing $Q$. However, near onset it is enhanced: The Dufour–induced destabilization of the conductive state shifts the convective onset to lower values of $r$. The reduction of $\mathbf{X}^2$ at large $r$ is nearly proportional to $Q$. This can be checked by an expansion for $r \to \infty$, which already holds for $r \gtrsim 3$, where all curves $\mathbf{X}^2(r)$ tend asymptotically to straight lines.

The most conspicuous change in the SOC bifurcation behaviour with increasing Dufour effect is the gradual change from a strongly backwards bifurcation ($Q = 0$ in Fig. 3) via a tricritical one to a forwards bifurcation. This behaviour is documented in a more global manner in Fig. 4: The bifurcation of SOCs with $\hat{k} = 1$ is forwards (backwards) in the shaded (white) $L$-$\psi$-region to the right (left) of the thick full curve of tricritical bifurcations. The shaded region strongly grows with increasing $Q$ on cost of the white one. At the thin solid line of Fig. 4, the bifurcation threshold has moved to $r_{stat} = \infty$. So, to the left of it, the lower branch of $\mathbf{X}^2(r)$ is disconnected from the $\mathbf{X} = 0$ solution. In this regime, convection branches out of the conductive state at a finite negative $r_{stat}$, i. e., for heating from above.

Note that the tricritical bifurcation line at, e.g., $Q = 15$ in Fig. 4, is non monotonous — the Dufour-induced appearence of strongly nonlinear $L$-$\psi$ variations in convective and in stability [14] properties is not surprising in view of the fact that the Dufour effect enters via $LQ\psi$ and $LQ\psi^2$ into the field equations. Thus, when decreasing here $\psi$ from zero, e.g., along the dotted line in Fig. 4, one can observe the succesion $f \to t \to b \to t \to f \to t \to b$ of forwards ($f$), tricritical ($t$), and backwards ($b$) bifurcations.

In Fig. 3 we have seen that the flow intensity $\mathbf{X}^2$ decreases with increasing Dufour effect $Q$ or with decreasing heating rate, $r$. In Fig. 5 we show that the structural changes resulting from either increasing $Q$ or decreasing $r$ are in fact the same: The simultaneous agreement in the structure of all fields — temperature, velocity, and concentration — is almost quantitative for the two combinations $Q = 10$, $r = 2.5$ and $Q = 0$, $r = 2$ shown in Fig. 5 and could be made perfect by judiciously choosing $r$–$Q$–combinations. Compare also the lateral profiles of these fields at midheight of the fluid layer as shown in the second row of Fig. 5. Thus, increasing both, $Q$ and $r$, appropriately does not change the SOC state.

In the second row of Fig. 6 we show, for various Soret coupling strengths $\psi$, the influence of the Dufour effect on the $r$–variation of the model's Nusselt number (2.18,4.9) in SOC states. The influence of the Dufour effect on the Nusselt number is similar to that one on the flow intensity — cf. Fig. 3. First of all, the onset of convection is shifted with increasing $Q$ to a smaller Rayleigh number. Simultaneously, at larger $r$, say above $r = 2$, the vertical heat current decrases with increasing $Q$. Both effects combined flatten the bifurcation curve of $N - 1$ versus $r$ and/or change the backwards bifurcation topology at sufficiently negative $\psi$ into a forwards one. For example, in the extreme case of $\psi = -0.5$, where for $Q = 0$ the lower bifurcation branch (cf. dots in Fig. 6) is disconnected from the conductive state, already a Dufour effect of size $Q = 5$ brings down the onset $r_{stat}$ to about 3.47. Increasing $Q$ further the bifurcation becomes forwards even for this strong Soret coupling $\psi = -0.5$.

In the third row of Fig. 6, the graphs of $M$ (2.19,4.10) versus $r$ show how the Dufour effect influences the mean square variation of the concentration field in the SOC states of the model. Remember that $M$ is defined to be 1 in the conductive state. Furthermore, the better the convective mixing of the fluid the smaller are concentration variations and with it $M$. So we see in Fig. 6 that the Dufour–induced reduction of the convective flow intensity and of the Nusselt number at larger $r$ is accompanied by a reduction of the convective mixing: For larger $r$, the parameter $M$ increases with increasing $Q$. Roughly and qualitatively speaking the bifurcation behaviour of the flow intensity, $w_{max}^2$, and of the convective heat current, $N - 1$, are similar to $1 - M$ which measures the degree of convective mixing.

### 2. Traveling wave convection

Our Galerkin model has nonlinear convective solutions in the form of harmonic waves of constant amplitude traveling with constant phase speed $v_p = \omega/k$ either to the left or to the right. The complex Galerkin modes (4.4) for this TW solution have the form

$$\mathbf{X}(t) = |\mathbf{X}| e^{i\omega t} \quad (4.19a)$$
$$\mathbf{Y}(t) = |\mathbf{Y}| e^{i(\omega t + \alpha)} \quad (4.19b)$$
$$\mathbf{U}(t) = |\mathbf{U}| e^{i(\omega t + \beta)} \quad (4.19c)$$

with constant moduli $|\mathbf{X}|$, $|\mathbf{Y}|$, and $|\mathbf{U}|$ and phase differences $\alpha$, $\beta$. The real modes $Z$ and $V$ (4.5) are time independent as well. The hydrodynamic fields follow from (4.1).

From the mode equations (4.2) one finds that both $|\mathbf{X}|^2$ and the squared frequency $\omega^2(k)$ of the TW solution vary linearly with the distance from the oscillatory threshold at $r_{osc}(k)$:

$$|\mathbf{X}|^2 = s_{TW}(r - r_{osc}) \quad (4.20a)$$
$$\omega^2 = \omega_H^2 + f_{TW}(r - r_{osc}) \quad . \quad (4.20b)$$

The threshold values $r_{osc}(k)$ and $\omega_H(k)$ as well as the slopes $s_{TW}$ and $f_{TW}$ of the bifurcating TW solution branch depend on $L$, $Q$, $\psi$, and $\sigma$ but the $r$ dependence is always linear, i.e., of the same form found for $Q = 0$ and liquid mixture parameters with FSI boundary conditions [21,27]. The subcritically (supercritically) bifurcating TWs with $s_{TW} < 0$ ($s_{TW} > 0$) are unstable (stable, at least close to onset). At

$$r^* = r_{osc} - \frac{\omega_H^2}{f_{TW}} \quad (4.21)$$

the TW solution ends by merging with zero frequency into the SOC solution (4.18).

Fig. 7 shows existence boundaries of the TW solutions (stable or unstable). For $\sigma = 1$ the model has TW solutions with $k = k_c^0$ in the $Q$–$\psi$–region below the *thick* lines for various Lewis numbers as indicated by the line type. These existence boundaries are determined by the merging of the oscillatory and stationary bifurcation threshold, $r_{osc}(k_c^0) = r_{stat}(k_c^0)$, for fixed wave number $k_c^0$, i.e., by the codimension–two (CT) condition with vanishing Hopf frequency $\omega_H(k_c^0) = 0$. Note that the existence range of TWs with the critical oscillatory wave number $k_{osc}^c$ is somewhat wider than the one for TWs with $k_c^0$ as can be inferred also from Fig. 2. The boundary curves of Fig. 7 show that the $\psi$–range in which TWs can be found is shifted to more negative values when (i) the Lewis number increases, i.e., when concentration diffusion becomes more efficient or (ii) when the Dufour coupling increases. For $L \lesssim 0.6$ the $\psi$–range of TW existence range splits into two pieces for sufficiently large $Q$. A similar feature was found in the linear analysis of the regions with non vanishing Hopf frequency (Fig. 6 of [14]).

The TWs bifurcate supercritically (subcritically) out of the conductive state for the parameters in the shaded (white) regions of Fig. 7 below the respective *thick* curves. For parameters on the *thin*, vertically oriented lines, the TW bifurcation is tricritical. Obviously, forwards bifurcating TWs disappear pretty soon with growing Dufour number while backwards bifurcating ones can be seen for larger $Q$, provided the Soret coupling is sufficiently negative. The width of the shaded existence range in Fig. 7 of *supercritical* TWs, i.e., the distance $|\psi_{CT} - \psi_t|$ between CT and tricritical separation ratio measured, e.g., for $Q = 0$ and $\sigma = 1$ is largest at $L = 0.736$. This distance decreases to zero for $L \to 0$.

Our model also shows for *positive* $\psi$ TW solutions like the FSI model for $Q = 0$ [21,27]. These TWs at $\psi > 0$

branch in a secondary bifurcation at $r^*$ out of the SOC state whereas for $\psi < 0$ TWs bifurcate at the oscillatory threshold $r_{osc}$ out of the conductive state and end at $r^*$ in the SOC state. We should like to mention that numerical simulations of *liquid* mixtures gave no indication for the existence of TWs at positive $\psi$ (cf. [16] for a short discussion) which suggests that they result from the mode truncation of the model. For *gas* mixtures numerical results are not available. Furthermore, the model TW states at positive $\psi$ appear in gases at large $r$ (e.g., $r \gtrsim 10$ for $Q = 10$, $\psi = L = \sigma = \hat{k} = 1$) where with the mode amplitudes being large the model's truncation approximation is presumably not justified. And, finally, these states occur for relatively large positive separation ratios that might be inaccessible experimentally.

## V. COMPARISON WITH RESULTS FROM THE FULL FIELD EQUATIONS

In this section we compare linear as well as nonlinear properties of our eight–mode Galerkin approximation with results obtained from the full hydrodynamic field equations. In Subsection V A critical linear model properties are compared with an analytical exact analysis of the stationary instability of Sec.III and with results obtained numerically with a shooting method for the oscillatory instability [14]. In Subsection V B we compare nonlinear convective properties of the model with results from solving the field equations numerically with a finite-differences method.

### A. Linear properties

First we compare the numerical exact critical wave numbers, stability thresholds, and Hopf frequencies of Fig. 7 of [14] with Fig. 2 showing in an analogous way the results of our model. Our approximation of the critical wave number of the stationary patterns is nearly perfect with errors less than 1 % in the whole $Q$- and $\psi$-range investigated here. The stationary stability threshold $r^c_{stat}$ itself is approximated with the same quality over the whole $\psi$-interval for $Q \lesssim 10$ and for all investigated values of $Q$ for $\psi \gtrsim -0.3$. The build up of a local minimum and maximum in $r^c_{stat}(\psi)$ occuring for $Q \gtrsim 10$ in our model starts for smaller values of $Q$ in the numerical threshold [14]. This reflects small errors for moderate $Q$ and strong $\psi \lesssim -0.6$. Our model reproduces the $\psi$-value where $r^c_{stat}(\psi)$ diverges with an accuracy of about 5 % since the model's stability curve does not show the the proper scaling behaviour (see Sec.III) resulting from the full field equations.

The errors of our model for the oscillatory stability analysis are largely caused by a shifted $\psi$-dependence. This shift from the model values to those of the shooting analysis increases with stronger Dufour effect from 0 for $Q = 0$ to $-0.3$ for $Q = 20$ in the direction of the negative $\psi$-axis. After this transformation the model describes $r^c_{osc}$ as well as the Hopf frequency with small errors. Even the slight kink in the curve $\omega_H(\psi)$ for $Q = 5$ in the neighbourhood of $\psi \simeq -0.5$ is reproduced. In addition, our model provides good approximations for $k^c_{osc}$ if we restrict ourselves to values of $Q$ less than 5. For stronger Dufour effect it cannot produce the strong $\psi$-dependence seen in the shooting analysis, but it shows the main two qualitative effects of increasing $Q$ on the critical oscillatory wave numbers: a decreasing value of $k^c_{osc}$ and a building up of a minimum in the curve $k^c_{osc}(\psi)$.

To conclude, our NSI–Galerkin–approximation is very good as far as the stationary stability analysis is concerned. The errors in its oscillatory part are small and increase with $Q$. But they are less relevant, say, from an experimental point of view because they occur mostly in a parameter range where the first instability is stationary.

To get further insight into the reasons of the errors of our model we compare in Fig. 8 the vertical profiles of the critical modes for the stationary and the oscillatory instability, respectively, for $Q = 5$ as a representative example. Exact results are shown by full lines and the model's results by dashed lines. Because of the mirror symmetry of the critical modes at the mid plane $z = 0$, the modulus and phase profiles are displayed only in one half of the layer.

Consider first the profiles of the stationary critical modes (left row in Fig. 8). Those of the $\theta$-field contain with increasing $Q$ and $|\psi|$ a small admixture of higher vertical modes beyond the $\cos \pi z$ of the model. The critical velocity field is fitted very well by the first Chandrasekhar function excepted for a small error in its central maximum which increases with $Q$. The most important deviation occurs for the critical $\zeta$-field: The approximation $\zeta = const$ ignores the $z$-variation of the exact profile. But the constant lies for relevant fluid parameters within the range of variation of the exact result.

In the right row of Fig. 8 we compare moduli and phases of the critical oscillatory modes of the model with the corresponding exact results. Also here, like for the stationary instability, the vertical profiles of the critical eigenfunctions of velocity and temperature resulting from the model agree quite well with the exact ones in modulus and phase. However, the model phases are constant and typically larger than the $z$-dependent exact phases. Again the largest deviations, in modulus and phase, occur for the critical $\zeta$-field.

### B. Nonlinear Properties

To test the quality of the model's predictions for the nonlinear convective states we have performed some selected numerical simulations of the full hydrodynamic field equations (2.4) with a MAC/SOLA finite-differences method. This code has been employed suc-cessfully for binary liquid mixtures [16].

### 1. Traveling wave convection

For gas mixtures $L = 1 = \sigma$ and $0 \leq Q \leq 20$ we found in the full–field simulations no stable, *large–amplitude* TW solutions that have bifurcated subcritically in agreement with the model. Note that the existence range of nonlinear stable TWs on the upper TW solution branch in binary *liquid* mixtures at $Q = 0$ rapidly shrinks to zero as $L$ approaches 1 from below (Fig.14 in [16]). For *gas* mixtures with $L = 1 = \sigma$ the subcritically bifurcated TW solution branch ends on the SOC branch without having formed a saddle. Thus, this TW branch remains unstable all the way from the bifurcation threshold $r_{osc}$ to the end point $r^*$ on the SOC solution branch. This result is supported by a recent test calculation with a 267–mode Galerkin expansion. In it the number of $\theta$– and $\zeta$–field modes was large enough to reproduce any structural details of the finite–difference solution of the field equations.

The model shows in a small $Q$-$\psi$-region — cf the shaded areas of Fig. 7 — forwards bifurcating TWs (e.g., for $Q = 0$, $L = 1 = \sigma$, and $k = k_c^0$ between the tricritical value $\psi_t = -0.3812$ and the codimension–two value $\psi_{CT} = -0.2174$). However, the $r$-interval $(r_{osc}, r^*)$ where these supercritical TW solutions appear is very small — less than 1% of the current heating rate $r$ for $Q = 0$ — so that the initial slope of the bifurcating TW branch is always large. On the other hand, in the 267–mode Galerkin calculations, done for $Q = 0$, $\sigma = 1$, we found only backwards bifurcating TW solutions for $L = 1$ while runs for $L = \frac{1}{2}$ did show a small $\psi$–range near the CT point with supercritical TWs. So the domain boundaries between supercritical and subcritical TW states in the $L$-$\sigma$-$\psi$-$Q$ parameter space are not reproduced quantitatively by our 8–mode model.

### 2. Stationary convection

In Fig. 6 we compare the bifurcation properties or our model SOC solutions (4.18) with finite–differences simulations of the full field equations. We show the Nusselt number $N$ and the mixing parameter $M$ versus $r$ for four Dufour numbers ($Q = 0, 5, 10, 20$) at four different separation ratios — note the different abscissa and ordinates scales. Fig. 6 shows that our model reproduces the SOC bifurcation diagrams of $N$ and $M$ including the Dufour–induced trend towards a supercritical bifurcation topology with less convection and less mixing not only qualitatively but also semiquantitatively.

For a more detailed comparison of the SOC field structure of the model with that one of the numerically simulated solution we show in Fig. 9 vertical profiles of the first lateral Fourier modes $n = 0, 1, 2$. The velocity field is well approximated by our first lateral Fourier mode. Higher modes not contained in our model carry at most 5% of the convection, e.g., for the strongly nonlinear state at $r = 3.5$. Of course, their relative contribution increases with $r$. At $r = 2$, slightly above the saddle of the SOC-solution, our model reproduces the exact results for the temperature field with the same accuracy as for the velocity field. Here, increasing the Rayleigh number leads to a building up of a plateau in the first mode. This structure cannot be reproduced by our model since it takes only a $\cos \pi z$ into account. On the other hand, the model approximates the zeroth lateral mode in a nearly perfect manner. Higher modes in the temperature field contribute with about 10%. The largest errors occur in the concentration field. While they are still acceptable at $r = 2$ there are structural differences to be seen in the strongly nonlinear state at $r = 3.5$. There the exact profile shows a local maximum structure caused by the combination of the terms 1, $\cos \pi z$ and $\cos 2\pi z$. The latter one is absent in our approximation. In contrast to this first lateral mode our model reproduces the zeroth lateral mode of the concentration with the accuracy obtained for the velocity and temperature field: Apart from quantitative errors (of about 15% for $r = 3.5$) the model also reproduces the inversion of the central concentration gradient as the numerics do. For nonlinear stationary convection in $\psi < 0$–mixtures we can therefore state that the stable layering of the concentration in the purely heat conducting state is inverted in nonlinear states. This leads to a strong mixing of the fluid in the bulk of the fluid near $z = 0$. The profiles of our model show similar qualities and defects for positive $\psi$. There the nonlinear states invert the unstable layering of the conductive state into a stable layering. We finally stress that, e.g., the third lateral concentration mode that is not shown in Fig. 9 for the sake of clarity contributes with 20% relative to the first mode.

Altogehter the model predicts the vertical profiles of the first lateral modes reasonably well near the saddle of the SOC's. There, even the concentration field is approximated in an acceptable manner. In strongly nonlinear states we can approximate the vertical velocity field and that of the temperature with a satisfying quality.

## VI. SUMMARY

We have investigated the influence of the Dufour coupling, i. e., the effect that concentration gradients drive heat currents or change the temperature field on convection in binary fluid layers heated from below. The Dufour effect changes the temperature field equation "diagonally" via the term $LQ\psi^2 \nabla^2 T$ that enhances heat diffusion and "off–diagonally" via the term $-LQ\psi \nabla^2 C$ that reflects concentration-induced temperature variations. Thus the largest effects are seen for large $L$ and large $| \psi |$. We focussed our main interest on *gas* mix-tures with Lewis and Prandtl numbers around one. For Dufour numbers $0 \leq Q \lesssim 40$, i. e., in a range that seems to contain physically realistic gas parameters we have determined the influence of the Dufour effect, both on the linear stability behaviour of the quiescent conductive state and on various nonlinear convection properties in a range of experimentally relevant Soret couplings $\psi$. To that end we have used an eight–mode Galerkin approximation that describes convection in the form of straight parallel rolls subject to realistic vertical boundary conditions. Its nonlinear properties were compared with some selected numerical simulations of the full hydrodynamic field equations and the linear ones were compared with numerically or analytically exact stability analyses.

*LINEAR PROPERTIES* — The influence of the Dufour effect on the stability of the conductive state can be summarized — see also [14] — as follows: (1) It destabilizes (stabilizes) the conductive state against the growth of stationary (oscillatory) convection. (2) The critical curves $r^c_{stat}(\psi)$ and $r^c_{osc}(\psi)$ are shifted with increasing $Q$ towards more negative $\psi$. (3) Thus the $\psi$-range with stationary (oscillatory) instabilities grows (shrinks) with $Q$. (4) The Hopf frequency decreases with growing $Q$. These properties are well reproduced by our few–mode model. In agreement with [15] we found that the critical stationary wave number $k^c_{stat}(p)$ in the exact stability analysis is governed by only a single parameter $p$ that also determines the shape of the stationary marginal stability curve $\frac{1}{S}\tilde{R}_{stab}(k;p)$. Thus, $L$, $Q$, and $\psi$ enter only via the scaling combinations $p = \frac{\psi}{L(1+Q\psi^2)(1+\psi)}$ and $S = (1+Q\psi^2)(1+\psi)+\psi/L$. The exact value for which a stationary zero wave number instability occurs is found to be $p_0 = \frac{131}{34}$ and an expansion around this point is presented.

*NONLINEAR PROPERTIES* — To determine the influence of the Dufour effect on nonlinear convective properties we investigated in particular the bifurcation behaviour of the flow intensity, of the convective heat current $N-1$, and of the convective mixing of the concentration field $M$ as functions of $r$ for several $Q$.

*Traveling wave convection* — (1) The $\psi$-range in which TW solutions — stable or unstable — are present is shifted to more negative values when the Dufour coupling increases. (2) Forwards bifurcating TWs disappear already for smaller $Q$ while backwards bifurcating ones still exist for larger $Q$ when $\psi$ is sufficiently negative. (3) However, for $L = \sigma = 1$ these subcritically bifurcated TW solution branches do not develop a saddle before they merge with zero frequency with the SOC solution branch. Thus, *stable* TW states on an *upper* TW solution branch beyond a saddle that can be seen in liquid mixtures for $L = \mathcal{O}(10^{-2})$ und $\sigma = \mathcal{O}(10)$ were not found, neither in the model nor in the numerical simulations of the full field OBE's. The reason for their absence is the large Lewis number $L = \mathcal{O}(1)$ of gas mixtures.

*Stationary convection* — (1) The Dufour–induced destabilization of the conductive state against stationary perturbations shifts the onset of SOC to lower values of $r$. (2) The range of subcritical SOC bifurcations in the $L$-$\psi$ parameter plane shrinks upon increasing $Q$: (3) With growing Dufour coupling there is a gradual change from a strongly backwards SOC bifurcation — e. g. at large negative $\psi$ — via a tricritical one to a forwards one and the initial slope of a supercritical bifurcation curve decreases. (4) In addition, the Dufour effect reduces the convective intensity and with it the mixing of the concentration field at larger $r$ so that the bifurcation curves of $N - 1$, $w^2_{max}$, and $M$ as functions of $r$ become flatter. (5) Furthermore, structural changes in the convective fields resulting from either increasing $Q$ or decreasing $r$ are the same. (6) Comparison with numerical simulations of the full OBE's shows that the Galerkin approximation provides good results for stationary convection in gas mixtures also for SOC states that are well above onset not only for the Nusselt number but also for the convective mixing $M$. The reason is that the field expansion in trigonometric functions is quite effective since with $L = \mathcal{O}(1)$ the problem of concentration boundary layers is less severe than in liquids.

## ACKNOWLEDGMENTS

We want to thank P. Büchel and C. Jung for the numerical support concerning the simulation of the basic equations and the Shooting analysis for the oscillatory stability analysis, respectively. Discussions with W. Barten are gratefully acknowledged. This work was supported by Deutsche Forschungsgemeinschaft.

## APPENDIX A: FSI STABILITY ANALYSIS

Here we consider a FSI–mode truncation to study the influence of the Dufour effect on convection in binary gas mixtures with an idealized free–slip boundary condition for the velocity field but with impermeable plates using the ansatz

$$u_3 = \left[w_{11}e^{-ikx} + c.c.\right]\sqrt{2}\cos\pi z \quad \text{(A1a)}$$
$$\theta = \left[\theta_{11}e^{-ikx} + c.c.\right]\sqrt{2}\cos\pi z + \theta_{02}\sqrt{2}\sin 2\pi z \quad \text{(A1b)}$$
$$\zeta = \left[\zeta_{10}e^{-ikx} + c.c.\right] + \zeta_{01}\sqrt{2}\sin\pi z \quad \text{(A1c)}$$

Performing a stability analysis of the conductive state in the standard way we reproduce the stationary stability threshold $r_{stat}(k)$ and the critical wave number $k^c_{stat}$ obtained by Hort et al. [14]. But for the oscillatory instability we obtain results that differ for finite $Q$ from eqs. (3.17–3.19) of ref [14]. We get

$$r_{osc}(\hat{k}) = \frac{\hat{q}^6}{\hat{k}^2} \frac{\left[1 + \widehat{\mathcal{L}}\right]\left[(1+\sigma)\left(1+\frac{\widehat{\mathcal{L}}}{\sigma}\right) - \frac{\psi\widetilde{\psi}}{\sigma}\right]}{(1+\psi)(1+\sigma) - \frac{8\psi}{\pi^2}} \quad \text{(A2a)}$$

with a critical lateral wave number determined by a third-order polynomial in $\hat{k}_{osc}^{c\,2}$

$$\hat{k}_{osc}^{c\,6}\left[(1+\mathcal{L})\left(1+\frac{\mathcal{L}}{\sigma}\right)-\frac{\psi\tilde{\psi}}{\sigma}\frac{1+\mathcal{L}}{1+\sigma}\right]$$
$$+\hat{k}_{osc}^{c\,4}\left[3+\frac{\mathcal{L}^2}{\sigma}+2\mathcal{L}\left(1+\frac{1}{\sigma}\right)-\frac{\psi\tilde{\psi}}{\sigma}\frac{2+\mathcal{L}}{1+\sigma}\right]-4=0$$
(A2b)

and a Hopf frequency $\omega_H(\hat{k})$ given by

$$\omega_H^2 \frac{\tau^2}{\hat{q}^4} = -\widehat{\mathcal{L}}^2 - \frac{8\psi}{\pi^2}\frac{(1+\widehat{\mathcal{L}})(\sigma+\widehat{\mathcal{L}})}{(1+\psi)(1+\sigma)-\frac{8\psi}{\pi^2}}$$
$$+\psi\widehat{\tilde{\psi}}\frac{(1+\psi)(\widehat{\mathcal{L}}-\sigma)+\frac{8\psi}{\pi^2}}{(1+\psi)(1+\sigma)-\frac{8\psi}{\pi^2}}.\quad \text{(A2c)}$$

There we used the following abbreviations

$$R_c^0 = \frac{27}{4}\pi^4 \quad,\quad r = \frac{R}{R_c^0} \quad\text{(A3a)}$$

$$k_c^{0\,2} = \frac{1}{2}\pi^2 \quad,\quad \hat{k} = \frac{k}{k_c^0} \quad\text{(A3b)}$$

$$q_c^{0\,2} = \frac{3}{2}\pi^2 \quad,\quad \hat{q} = \frac{q}{q_c^0} \quad,\quad \tau = \frac{1}{q_c^{0\,2}} \quad\text{(A3c)}$$

$$\mathcal{L} = L(1+Q\psi^2) \quad,\quad \tilde{\psi} = \frac{8}{\pi^2}LQ\psi \quad\text{(A3d)}$$

$$\widehat{\mathcal{L}} = \frac{\hat{k}^2}{3\hat{q}^2}\mathcal{L} \quad,\quad \widehat{\tilde{\psi}} = \frac{\hat{k}^2}{3\hat{q}^2}\tilde{\psi} \quad.\quad\text{(A3e)}$$

In Fig. 10 we present our FSI results which show in contrast to Fig. 2 of [14] the characteristic Dufour–induced shrinking of the $\psi$–region with an oscillatory instability of the conductive state.

TABLE I. Critical stationary properties of binary fluid mixtures as a function of scaling variable $p$

| $p$ | $\widetilde{R}_c = \tau_c^3 k_c^4$ | $k_c$ | $\hat{w}_0$ | $10\hat{w}_1$ |
|---|---|---|---|---|
| -0.9 | 8353.80 | 6.67060 | -2.26232 | -0.641355+0.370865 i |
| -0.8 | 5189.08 | 5.67730 | -1.77924 | -0.599229+0.347371 i |
| -0.7 | 3831.61 | 5.01521 | -1.53865 | -0.478996+0.384074 i |
| -0.6 | 3099.23 | 4.54281 | -1.40431 | -0.327500+0.458674 i |
| -0.5 | 2645.13 | 4.18252 | -1.32071 | -0.162232+0.556742 i |
| -0.4 | 2336.96 | 3.89406 | -1.26455 | 0.00938873+0.669977 i |
| -0.3 | 2114.30 | 3.65476 | -1.22469 | 0.183866+0.793459 i |
| -0.2 | 1945.88 | 3.45082 | -1.19517 | 0.359434+0.924144 i |
| -0.1 | 1813.96 | 3.27336 | -1.17258 | 0.535157+1.06008 i |
| 0 | 1707.76 | 3.11633 | -1.15480 | 0.710531+1.19997 i |
| 0.1 | 1620.37 | 2.97546 | -1.14047 | 0.885289+1.34294 i |
| 0.2 | 1547.13 | 2.84765 | -1.12865 | 1.05930+1.48840 i |
| 0.3 | 1484.82 | 2.73056 | -1.11871 | 1.23250+1.63592 i |
| 0.4 | 1431.13 | 2.62240 | -1.11018 | 1.40489+1.78523 i |
| 0.5 | 1384.35 | 2.52177 | -1.10271 | 1.57649+1.93613 i |

FIG. 1. Dependence of stationary critical properties on the scaling variable $p$ (3.6). (a) Reduced critical wave number (solid line). The dashed line represents the expansion around $p_0 = 131/34$. (b) Reduced scaled stability threshold.

FIG. 2. Stability properties of a gas mixture ($L = 1, \sigma = 1$) vs. separation ratio $\psi$ for different Dufour numbers $Q$. The reduced stationary (solid line) and oscillatory (dashed line) stability thresholds $r^c_{stat}$ and $r^c_{osc}$, the corresponding reduced wave numbers $\hat{k}^c_{stat}$ and $\hat{k}^c_{osc}$, and the critical Hopf frequency $\omega_H$ result from the NSI model for which $R^0_c = 1728.38$ and $k^0_c = 3.098$.

FIG. 3. Square of the vertical velocity mode $\mathbf{X}$ of stationary convection vs reduced Rayleigh number $r$ for different values of the Dufour number $Q$. Parameters are $\psi = -0.25$, $L = 1$, $k = k^0_c$, and arbitrary $\sigma$. The long dashed line shows $\mathbf{X}^2 = r - 1$ in a pure fluid ($\psi = 0$). For comparison with experiments one should identify $\mathbf{X}^2$ with the flow intensity reduced by the pure fluid value at $r = 2$, i.e., $\mathbf{X}^2 = w^2_{max}/w^2_{max}(\psi = 0, r = 2)$.

FIG. 4. Onset behaviour of SOC with $k = k^0_c$ in the $L$–$\psi$ plane for different Dufour numbers. The bifurcation of flow intensity vs Rayleigh number is forwards (backwards) to the right (left) of the full thick curve of tricritical bifurcations. The threshold $r_{stat}$ has moved to $r_{stat} = \infty$ at the thin line. Below this curve the convective solution is disconnected from the conduction fixed point. In this parameter regime convection branches out of the conductive state for heating from above, $r < 0$.

FIG. 5. Structural properties of SOC states in the $x$–$z$ plane perpendicular to the roll axes. Black implies low field values and white high ones. The second row shows the lateral profiles of $w$ (solid line), $20\theta$ (dashed line), and $100c$ (dot–dashed line) at mid height of the fluid layer. Increasing $Q$ or decreasing $r$ leads to structurally similar fields — increasing both appropriately does not cause changes. Parameters are $\psi = -0.25$, $L = 1$, and $\hat{k} = 1$.

FIG. 6. Nusselt number $N$ and mixing parameter $M$ of stationary convection versus Rayleigh number $r$ resulting from our model and from numerical simulations of the full field equations with $L = \sigma = 1$, $k = k^0_c$. Within a column the Soret coupling $\psi$ has the value indicated in the figure. Curves for different Dufour numbers $Q$ are identified in the legend. Unstable branches (dotted lines) of subcriticallly bifurcating SOCs can be seen only in our model, namely for $\psi = -0.25$ and $\psi = -0.5$.

FIG. 7. Existence boundaries of TW solutions (stable or unstable) of the model for $\sigma = 1$, $k = k^0_c$. TWs exist below the thick labelled curves. For $k = k^c_{osc}$ the existence range is larger as can be inferred from Fig. 2. The shaded (white) regions show the range of supercritically (subcritically) bifurcating TWs.

FIG. 8. Vertical profiles of the linear critical Fourier modes. Full lines refer to the exact result obtained analytically (see Sec.III) or numerically with a Shooting method. Dashed lines represent the model. Because of the mirror symmetry of the critical modes at the mid plane, $z = 0$, only one half of the layer is shown. Left column contains the moduli at the stationary instability for $\psi = -0.3$. Right column shows the moduli and phases at the oscillatory instability for $\psi = -0.4$. The vertical average of $\varphi_w(z)$ has been assigned to the phase angle zero. The normalization is always such that $|\theta| = R^0_c$ at $z = 0$. Parameters are $\sigma = 1 = L$ and $Q = 5$.

FIG. 9. Vertical profiles of the lateral Fourier modes $n = 0$ (circles and full lines), $n = 1$ (squares and dashed lines), and $n = 2$ (triangles) for stationary convection. Symbols refer to finite–difference numerical simulations of the full field equations and lines represent our model. In the last two rows $\delta C$ denotes the deviation of the concentration from the global mean and the dotted line is the conductive profile of $\delta C$. Numerical profiles of, e.g., the $n = 3$ concentration mode contributing with about 20% of the first mode are not presented for the sake of clarity. Parameters are $\psi = -0.25$, $Q = 0$, $L = \sigma = \hat{k} = 1$.

FIG. 10. Stability properties of a gas mixture ($L = 1$, $\sigma = 1$) vs separation ratio $\psi$ for different Dufour numbers $Q$. The stationary (solid line) and oscillatory (dashed line) stability thresholds $r^c_{stat}$ and $r^c_{osc}$, the corresponding reduced wave numbers $\hat{k}^c_{stat}$ and $\hat{k}^c_{osc}$, and the critical Hopf frequency $\omega_H$ are determined approximately for FSI boundaries for which $R^0_c = \frac{27}{4}\pi^4$ and $k^0_c = \pi/\sqrt{2}$. This figure is meant s a corrigendum of Fig. 2 of [14].

**Fig. 1:**
Hollinger and Luecke
Influence of the Dufour effect on convection in binary gas mixtures
Phys.Rev.E

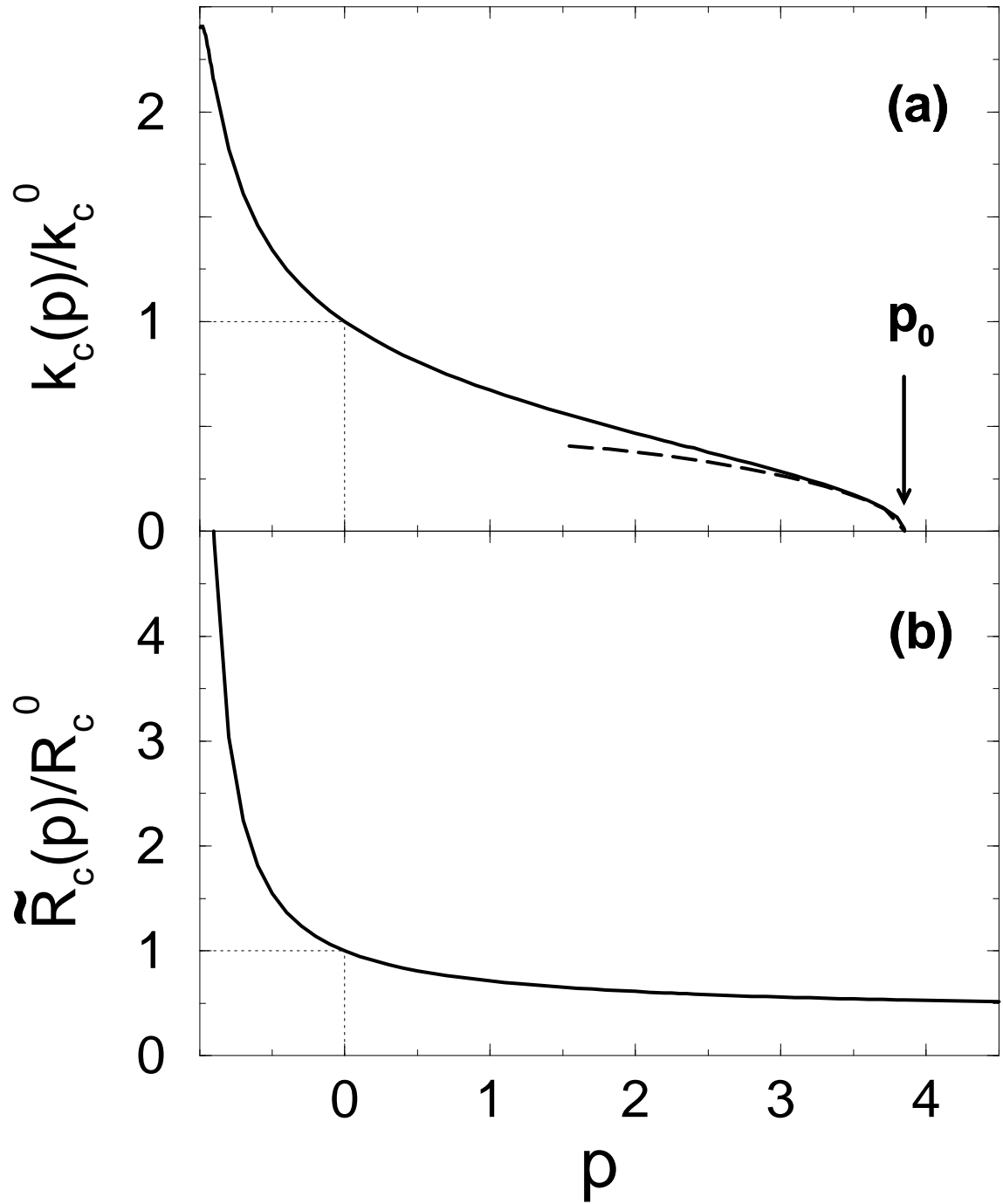

Fig.2:

Hollinger and Luecke
Influence of the Dufour effect on convection in binary gas mixtures
Phys.Rev.E

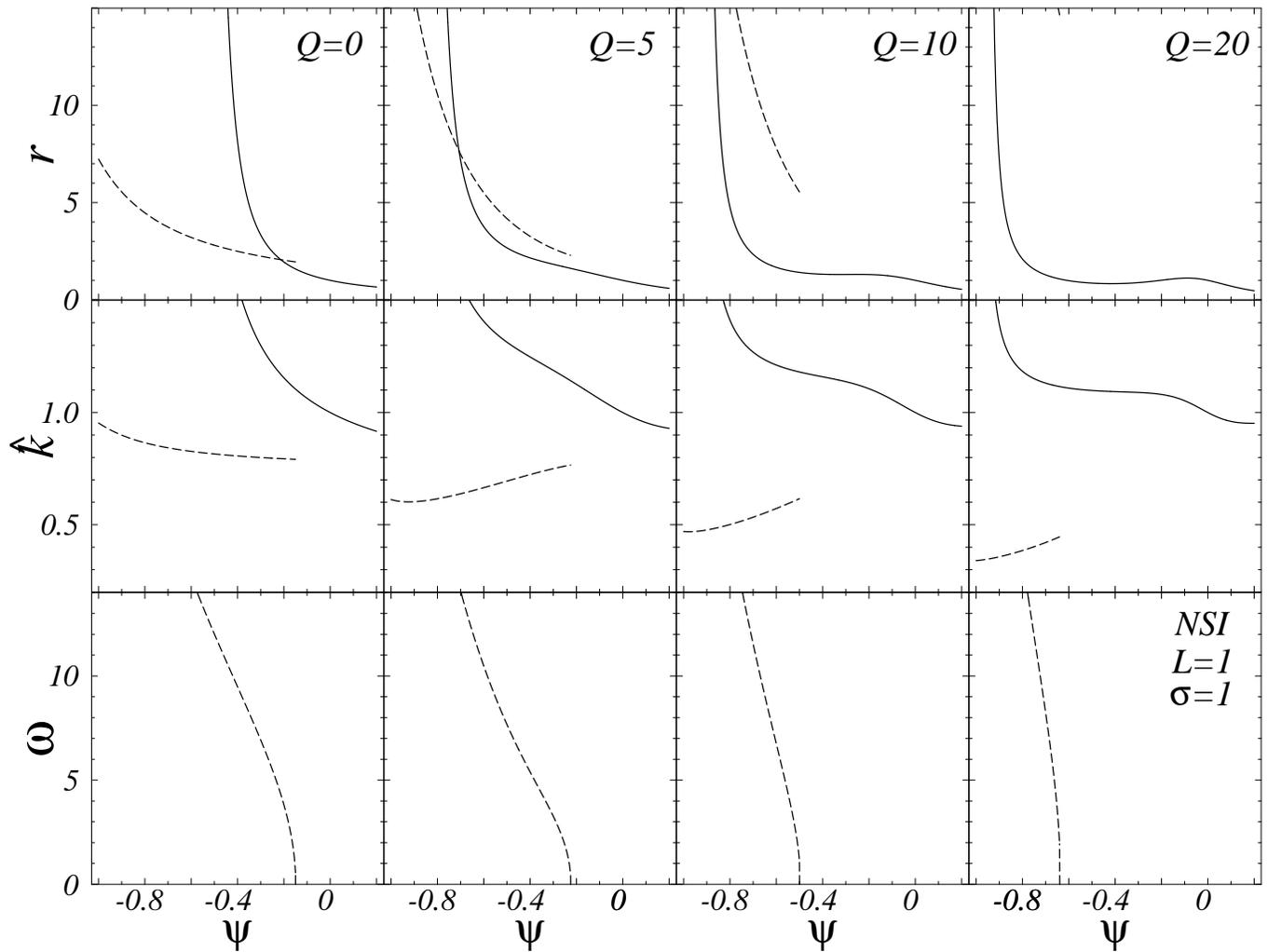

Fig.3:

Hollinger and Luecke
Influence of the Dufour effect on convection in binary gas mixtures
Phys.Rev.E

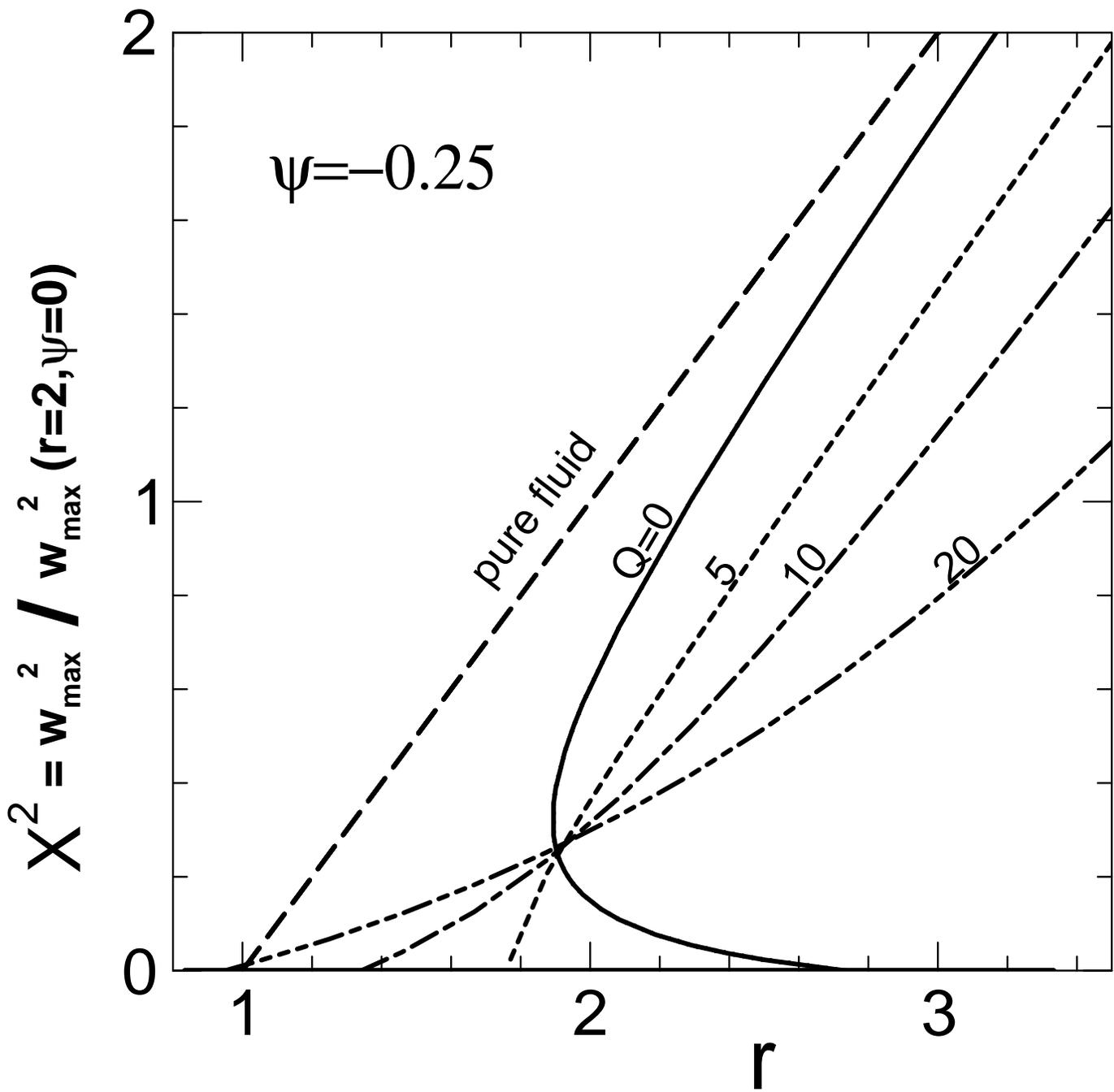

Fig.4:

Hollinger and Luecke
Influence of the Dufour effect on convection in binary gas mixtures
Phys.Rev.E

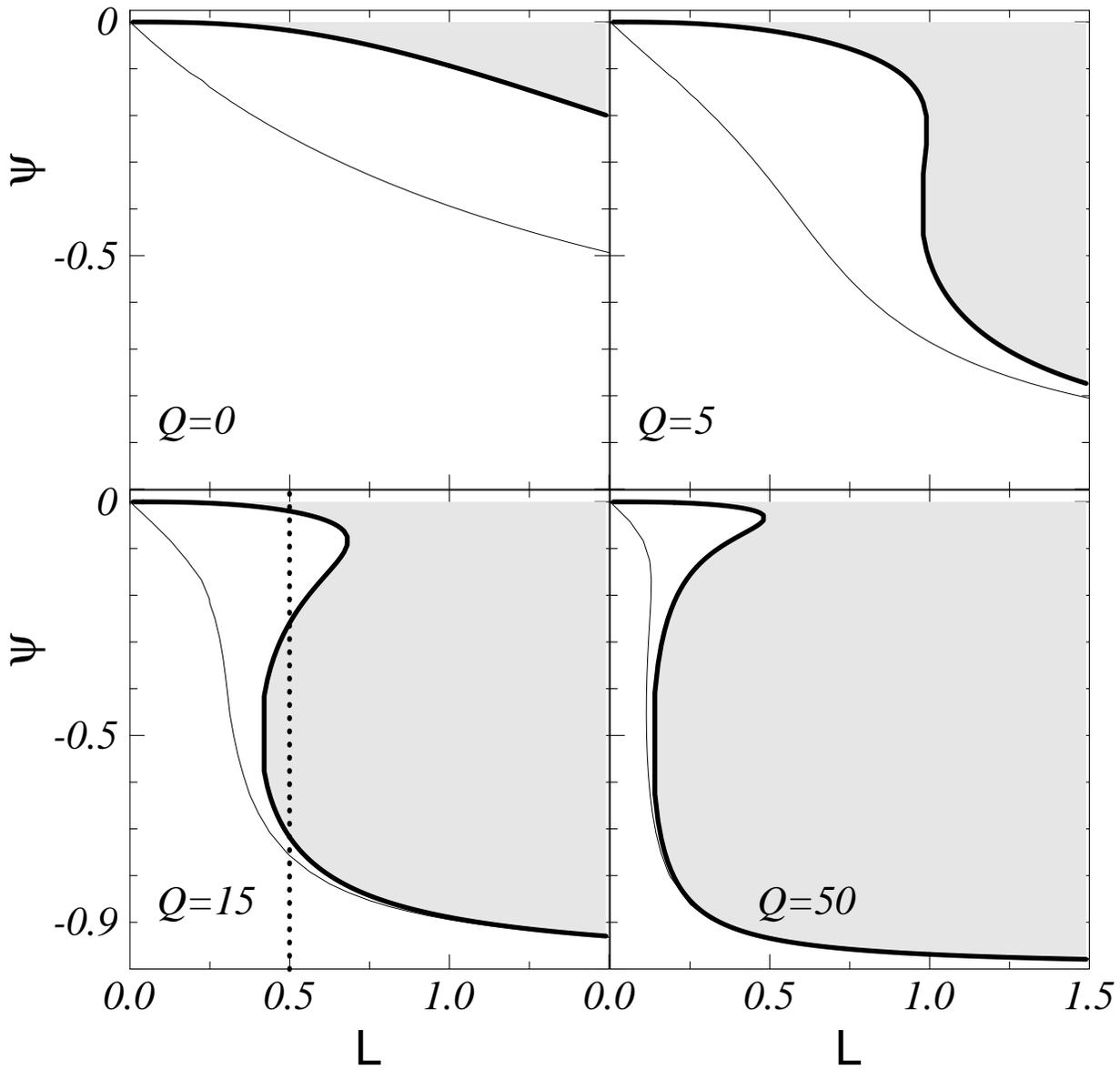

Fig.5:

Hollinger and Luecke
Influence of the Dufour effect on convection in binary gas mixtures
Phys.Rev.E

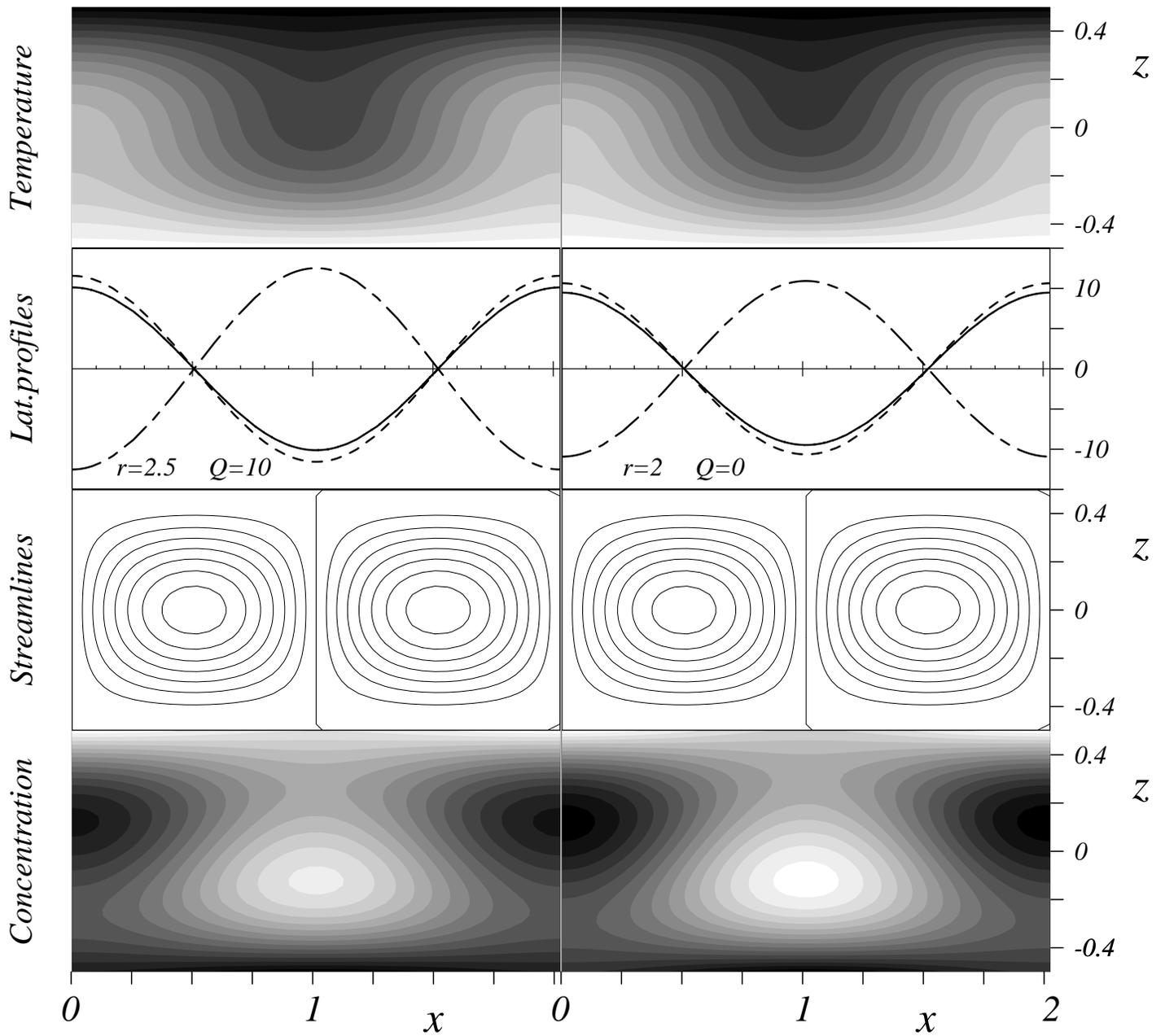

Fig.6:

Hollinger and Luecke
Influence of the Dufour effect on convection in binary gas mixtures
Phys.Rev.E

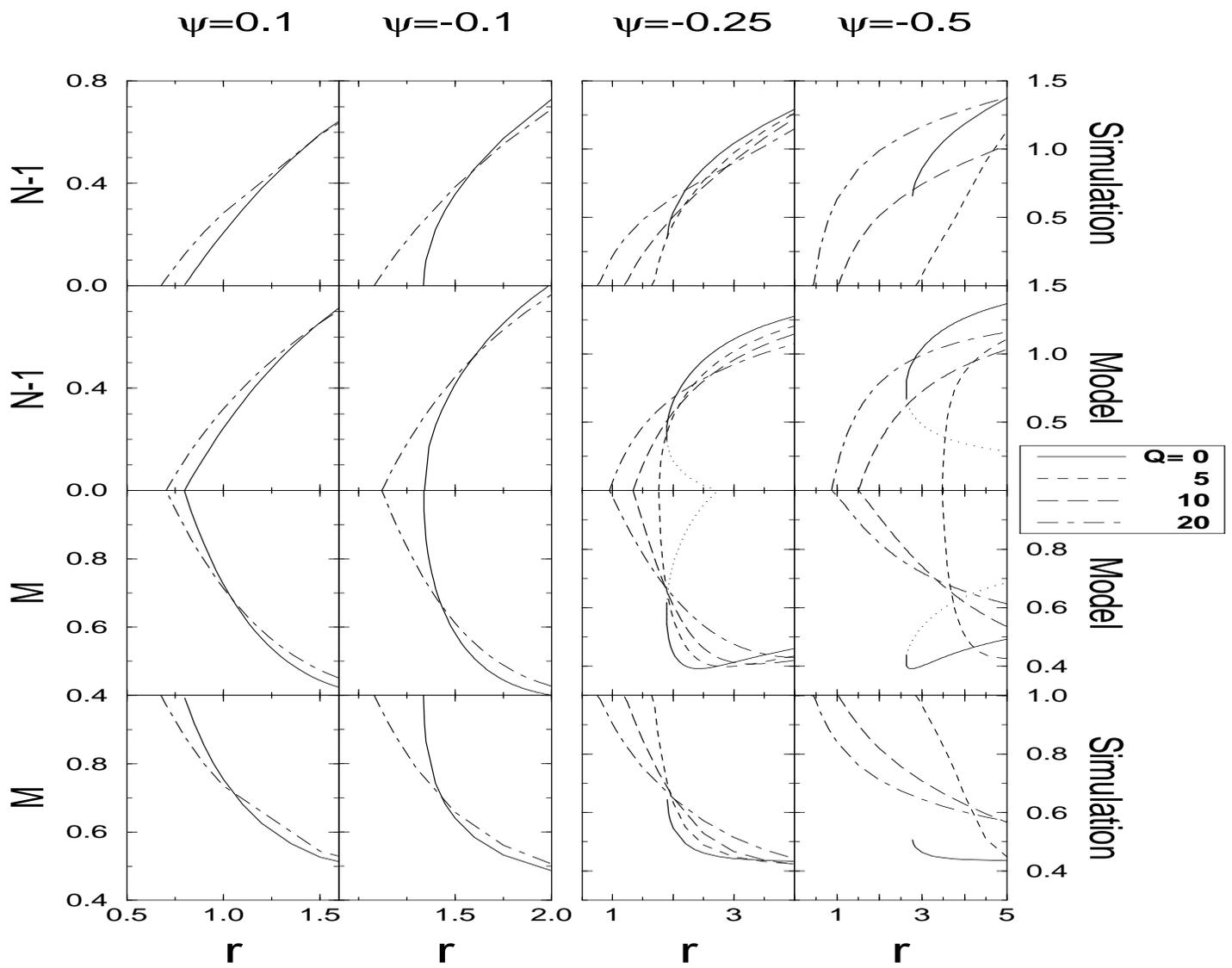

Fig.7:

Hollinger and Luecke
Influence of the Dufour effect on convection in binary gas mixtures
Phys.Rev.E

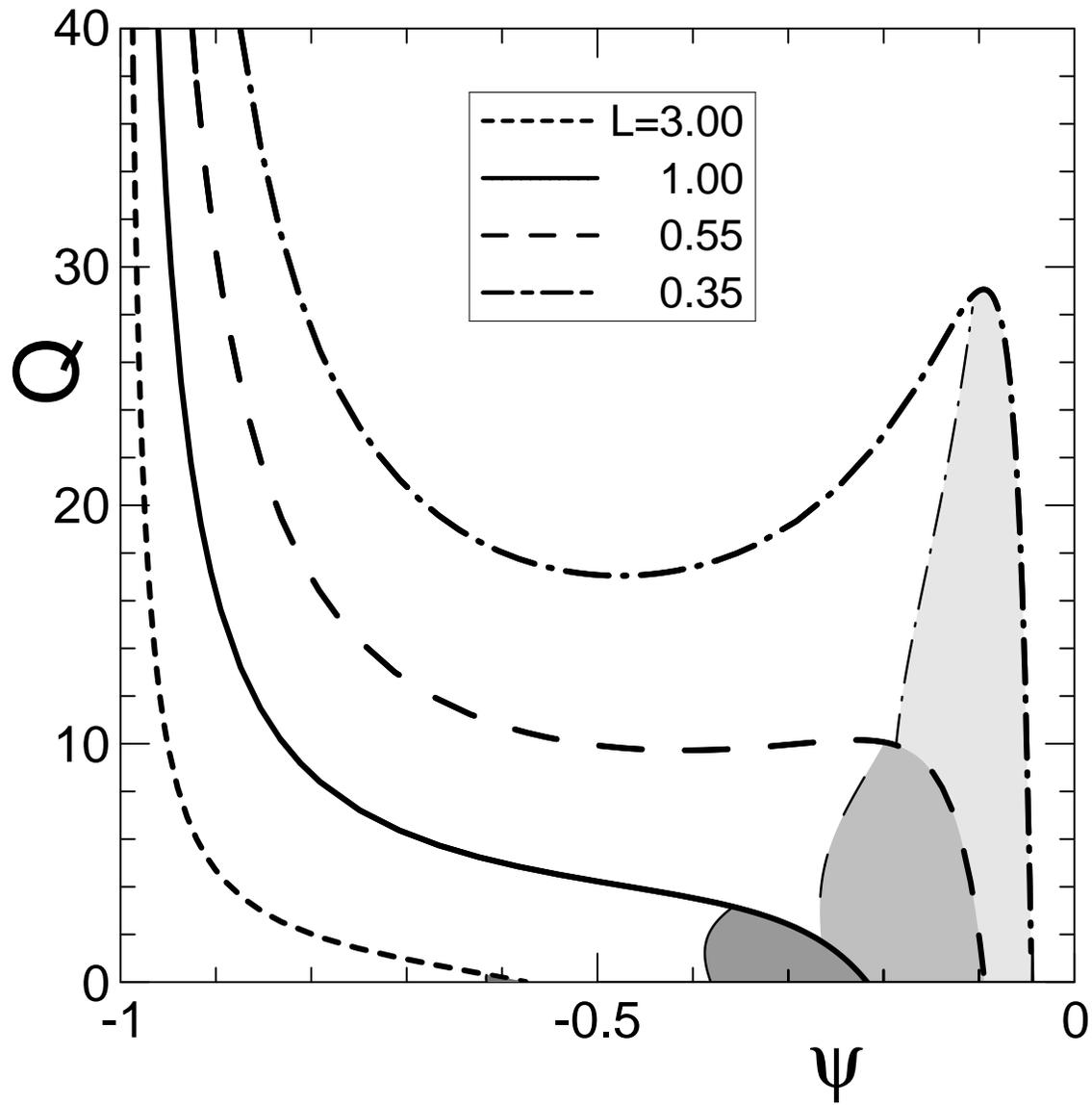

Fig.8:

Hollinger and Luecke
Influence of the Dufour effect on convection in binary gas mixtures
Phys.Rev.E

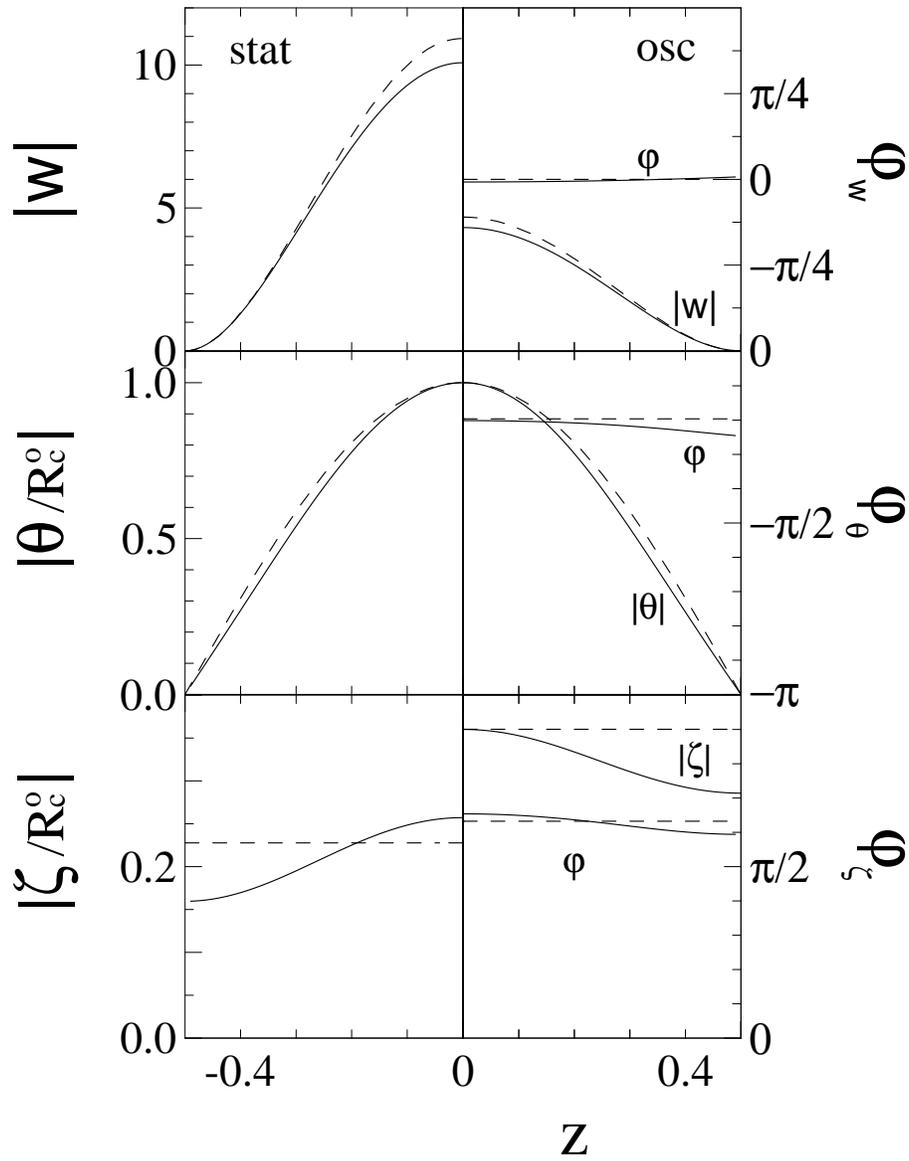

Fig.9:

Hollinger and Luecke
Influence of the Dufour effect on convection in binary gas mixtures
Phys.Rev.E

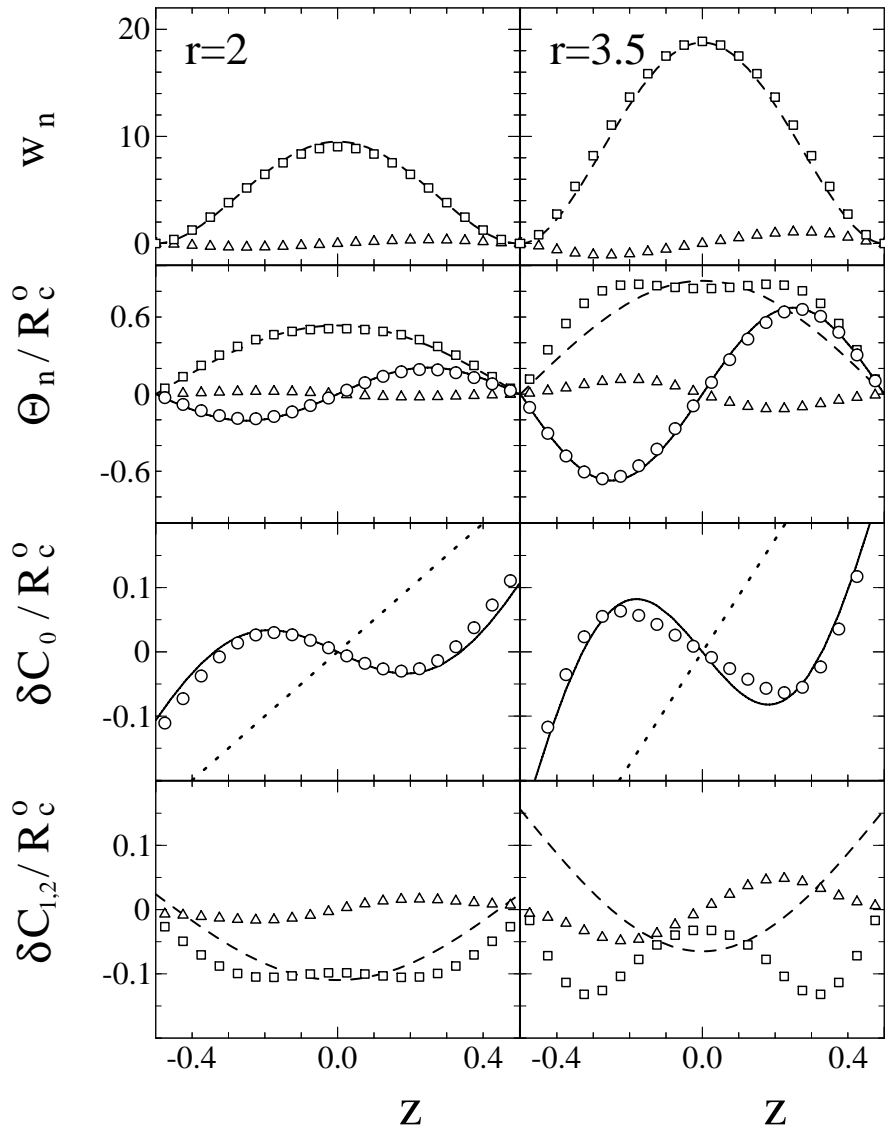

Fig.10:

Hollinger and Luecke
Influence of the Dufour effect on convection in binary gas mixtures
Phys.Rev.E

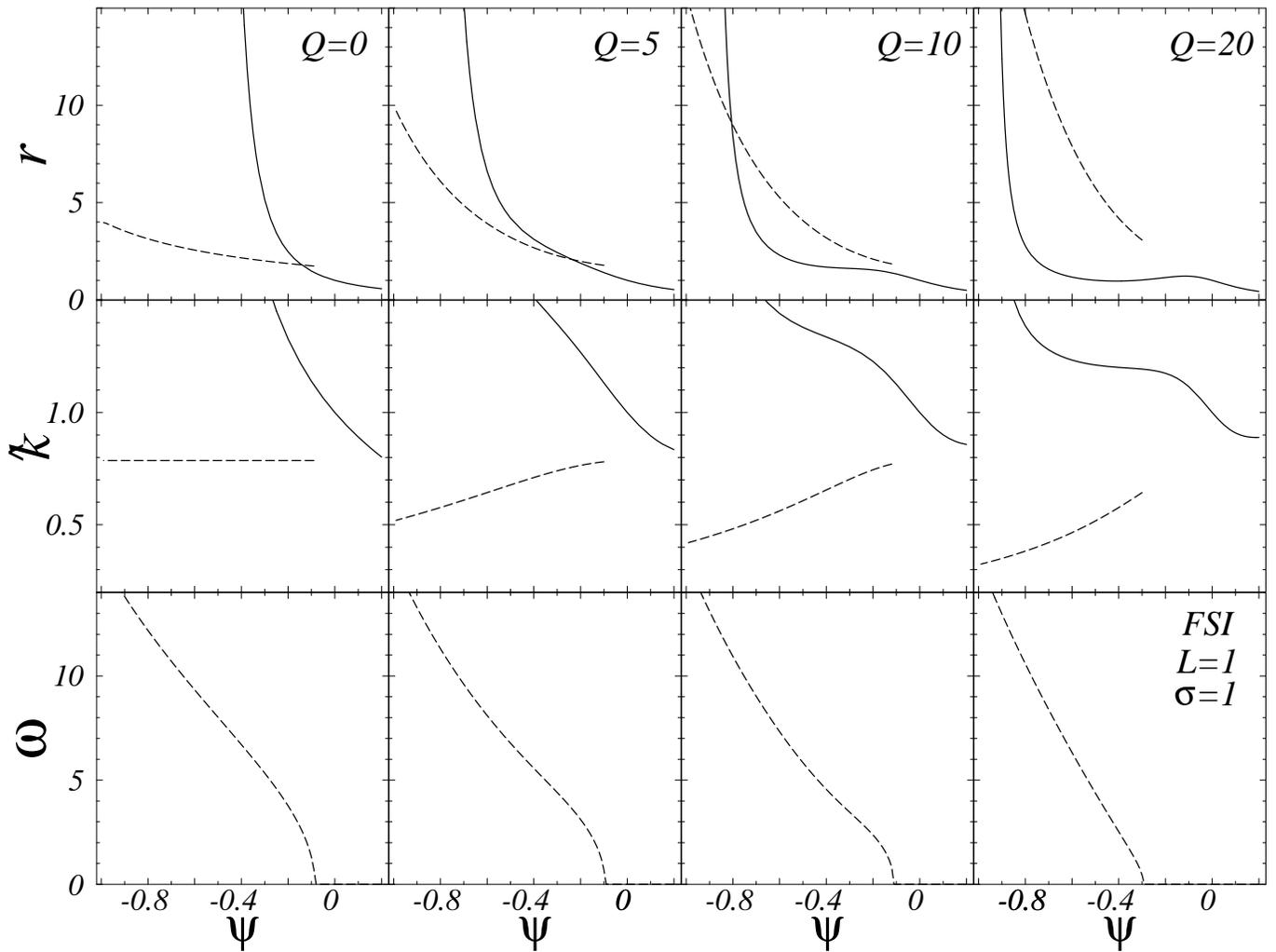